\documentclass[pre, aps,showpacs,supergroupedaddress,twocolumn,epsfig,amsmath,amssymb,eqsecnum]{revtex4}
\usepackage{epsfig,amsmath,amssymb,bm,epsf,graphics,psfrag,verbatim,subfigure,framed}
\usepackage[usenames,dvipsnames]{color}
\usepackage{bbm}
\usepackage{mathrsfs}
\usepackage{amsfonts}
\usepackage{color}
\usepackage{pbox}
\def\newblock{\hskip .11em plus .33em minus .07em}

\newcommand{\Limm}{\raisebox{0.5ex}{{$\displaystyle \lim_{\bm{r}^\prime\rightarrow\bm{r}}\;$}}}

\def\leftcoeff{\textit{c}}
\def\coupling{{g}}
\def\self{\Delta \varepsilon}
\def\bjerrum{b}
\def\gouy{\mu}
\def\debye{\ell_{{\rm DB}}}

\newcommand{\be}{\begin{equation}}
\newcommand{\ee}{\end{equation}}
\newcommand{\ba}{\begin{eqnarray}}
\newcommand{\ea}{\end{eqnarray}}
\newcommand{\bw}{\begin{widetext}}
\newcommand{\ew}{\end{widetext}}
\newcommand{\Tr}{{\rm{Tr}\,}}

\newcommand{\rv}{{\bm{r}}}
\newcommand{\xv}{{\bm{x}}}
\newcommand{\yv}{{\bm{y}}}

\newcommand{\kv}{{\bm{k}}}

\begin{document}
\title{The Correlation Potential of a Test Ion Near a Strongly Charged Plate}
\author{Bing-Sui Lu}
\email{binghermes@gmail.com}
\author{Xiangjun Xing}
\email{xxing@sjtu.edu.cn}
\affiliation{Department of Physics and Astronomy, and Institute of Natural Sciences, Shanghai Jiao Tong University, Shanghai, 200240, China}

\date{\today}
\pacs{82.70.Dd, 83.80.Hj, 82.45.Gj, 52.25.Kn}
\begin{abstract}
We analytically calculate the correlation potential of a test ion near a \emph{strongly} charged plate inside a dilute $m:-n$ electrolyte.  We do this by calculating the electrostatic Green's function in the presence of a nonlinear background potential, the latter having been obtained using the nonlinear Poisson-Boltzmann theory.  We consider the general case where the dielectric constants of the plate and the electrolyte are distinct.  The following generic results emerge from our analyses:
(1) If the distance to the plate $\Delta z$ is much larger than a Gouy-Chapman length, the plate surface will behave effectively as an infinitely charged surface, and the dielectric constant of the plate effectively plays no role.
(2) If $\Delta z$ is larger than a Gouy-Chapman length but shorter than a Debye length, the correlation potential can be interpreted in terms of an image charge that is three times larger than the source charge.  This behavior is independent of the valences of the ions.
(3) The Green's function vanishes inside the plate if the surface charge density is infinitely large; hence the electrostatic potential is constant there.  In this respect, a strongly charged plate behaves like a conductor plate.
(4) If $\Delta z$ is smaller than a Gouy-Chapman length, the correlation potential is dominated by the conventional image charge due to the dielectric discontinuity at the interface.
(5) If $\Delta z$ is larger than a Debye length, the leading order behavior of the correlation potential will depend on the valences of the ions in the electrolyte. Furthermore, inside an \emph{asymmetric} electrolyte, the correlation potential is \emph{singly} screened, i.e., it undergoes exponential decay with a decay width equal to the Debye length.
\end{abstract}

\maketitle
\section{Introduction}
\label{sec:intro}
The average electrostatic potential $\Phi(\rv)$ inside a $m:-n$ electrolyte satisfies the (exact) Poisson equation:
\be
\label{eq:Poisson}
- \epsilon \, \Delta \Phi(\rv) =
m q\, \rho_+ (\rv)
- n q\, \rho_- (\rv),
\ee
where $\Delta$ is the Laplacian,  $q = 1.6 \times 10^{-19} C$ is the electric charge of a monovalent ion, and $\rho_{\pm}(\rv)$ are the average number densities of positive ions (with charge $+mq$) and of negative ions (with charge $-nq$) respectively. Using statistical mechanics, it can be easily shown that the number density $\rho_{+}(\rv)$ is related to the {\em potential of mean force} $w_1(\rv,mq)$ of an ion of charge $mq$ via
\begin{subequations}
\label{rho-rho^0}
\ba
\rho_+(\rv) = \rho^0_+ \, e^{- \beta w_1(\rv,mq)},
\ea
where $\beta = 1/k_B T$ and $\rho^0_{+}$ is the number density in the bulk.  Similarly for negative ions we have
\be
\rho_-(\rv) = \rho^0_- \, e^{- \beta w_1(\rv,-nq)}.
\ee
\end{subequations}
The physical significance of $w_1(\rv, q)$ is the free energy cost of moving an ion $q$ from an infinite distance away to the position $\rv$ inside the electrolyte.

Let the electrolyte consist of $N$ mobile ions with charge strengths $q_i$ and positions $\xv_i$, $i = 1, 2, \ldots, N$.  The total Hamiltonian of the system is given by~\cite{footnote1}
\be
H_N = \sum_{i<j} q_i \, q_j\, v(\xv_i, \xv_j),
\label{H_N-def}
\ee
where $v(\xv, \yv)$ is the Coulomb potential at $\xv$ due to a monovalent positive ion at $\yv$.
Furthermore, let us insert a test ion of charge strength $k q$ (i.e., valence $k$) at the position $\rv$.  The potential at $\rv$ generated by all other ions $\{q_1, \ldots, q_N\}$ is then given by $\varphi(\rv)$:
\be
\varphi(\rv) = \sum_{i=1}^{N} q_i\, v(\rv,\xv_i).
\label{varphi-def}
\ee
The partition function $Z_N(\rv, q)$ of the electrolyte in the presence of the test ion can then be expressed as follows:
\ba
Z_N(\rv, k q) &=& \Tr_N \, e^{-\beta H_N -\beta k q\,\varphi(\rv)}
\nonumber\\
&\equiv&Z_N \frac{1}{Z_N} \Tr_N \, e^{-\beta H_N -\beta k q\,\varphi(\rv)}
\nonumber\\
&=& Z_N \, \left \langle  e^{ -\beta k q\,\varphi(\rv)} \right \rangle_N,
\label{ave-exponential}
\ea
where $\Tr_N$ denotes integration of all $N$ position vectors $\rv_i$ of mobile ions~\footnote{There is also a multiplicative factor coming with the integral.  It however does not affect our discussion. }, and $\langle \, \cdot \, \rangle_N$ denotes averaging over the Gibbs-Boltzmann distribution  $e^{-\beta H_N}$.   The potential of mean force $w_1(\rv, k q)$ is then related to the partition function $Z_N(\rv, k q)$ via
\be
w_1(\rv, k q) \equiv - T \ln \frac{Z_N(\rv, k q)}{Z_N(\infty, k q)}.
\label{w_1-definition}
\ee
 Note that $w_1(\rv, k q)$ is defined such that it vanishes as $|\rv|$ tends to infinity:
\be
w_1(\rv, k q) \rightarrow 0, \quad {\rm as} \quad |\rv| \rightarrow \infty.
\ee
Consequently $\rho^0_{\pm} $ in Eqs.~(\ref{rho-rho^0}) are indeed the ion number densities in the bulk.

The average of the exponential quantity in Eq.~(\ref{ave-exponential}) can be formally expressed as a cumulant series:
\be
 \left \langle  e^{ -\beta k q \,\varphi(\rv) } \right \rangle_N
 = \exp
 \sum_{j} \frac{(- \beta k q)^j}{j!}
  \langle \varphi(\rv)^j \rangle_c,
\label{cumulant-expansion}
\ee
where $\langle \varphi(\rv)^j \rangle_c$ is the $j$-th order cumulant of $\varphi(\rv)$:
\ba
\langle \varphi(\rv) \rangle_c &=& \langle \varphi(\rv) \rangle_N
= \Phi(\rv), \nonumber\\
\langle \varphi(\rv)^2 \rangle_c &=&
\langle \varphi(\rv)^2 \rangle_N
- \langle \varphi(\rv) \rangle_N^2, \\
\cdots &=& \cdots \nonumber
\ea
The cumulant expansion in Eq.~(\ref{cumulant-expansion}) can be formally understood as an expansion in terms of the valence $k$.

As the simplest approximation one keeps only the first cumulant in Eq.~(\ref{cumulant-expansion}):
\be
\langle e^{- \beta k q \varphi(\rv)} \rangle
\approx
e^{- \beta k q \langle \varphi(\rv) \rangle_N}
= e^{- \beta k q \Phi(\rv)}.
\nonumber
\ee
Combining this with Eqs.~(\ref{ave-exponential}) and (\ref{w_1-definition}), and choosing the convention that the average potential $\Phi$ vanishes in the bulk, i.e., $\lim_{\rv\rightarrow \infty} \Phi(\rv) = 0$, we find
\be
w_1(\rv, k q) = k q \, \Phi(\rv) + O(k^2).
\label{w_1-1}
\ee
which would be an exact equality if the potential $\varphi$ does not have any fluctuations.  Hence the approximation is essentially of mean field character.  Making this approximation for the density distributions of positive and negative ions, Eqs.~(\ref{rho-rho^0}), we arrive at the famous {\em Poisson-Boltzmann equation} (PBE):
\be
\label{eq:pbe}
- \epsilon \Delta \Phi(\rv) =
m q \rho^0_+  \, e^{- \beta m q \Phi(\rv)}
- n q \rho^0_- \, e^{\beta n  q \Phi(\rv)}.
\ee
Qualitatively speaking, PB theory hinges upon the assumption that ions are interacting with the local average potential $\Phi(\rv)$, instead of with other ions.  In the bulk,  Eq.~(\ref{eq:pbe}) reduces to $m\,\rho_+^0 - n\,\rho_-^0 = 0$, which can be understood as the condition of overall charge neutrality.

To obtain a better approximation, let us keep the second order cumulant in Eq.~(\ref{cumulant-expansion}).  For later convenience, let us also define the {\em reaction potential}
of a monovalent ion $\Upsilon(\rv,\rv')$ via
\be
\Upsilon(\rv,\rv') \equiv
- \beta q \langle \varphi(\rv)  \varphi(\rv')\rangle_c.
\label{Upsilon-def}
\ee
It is symmetric in two variables $\rv, \rv'$ by construction.
The potential of mean force of a test ion $k q$ is then given by
\be
w_1(\rv, k q) = k q \, \Phi(\rv)
+ \frac{1}{2}  k^2 q \, \delta \Upsilon(\rv, \rv)
+ O(k^3).
\label{w_1-2}
\ee
We shall call $\Upsilon(\rv,\rv)$ (i.e., the reaction potential for which $\rv'=\rv$) the {\em correlation potential}, and
\be
\delta \Upsilon(\rv, \rv) = \Upsilon(\rv, \rv)
- \lim_{\rv \rightarrow \infty} \Upsilon(\rv, \rv),
\ee
is then the correlation potential relative to its bulk value.  Comparing Eq.~(\ref{w_1-2}) with  Eq.~(\ref{w_1-1}), we see that the correlation
potential is responsible for the leading-order correction to the potential of mean force beyond Poisson-Boltzmann theory.


Substituting Eq.~(\ref{w_1-2}) back into Eqs.~(\ref{eq:Poisson}) and (\ref{rho-rho^0}), we arrive at a {\em fluctuation corrected Poisson-Boltzmann equation} (FCPBE):
\ba
\label{eq:mpbe}
- \epsilon \Delta \Phi(\rv) &=&
m q \rho^0_+ \,  e^{- \beta m q \Phi(\rv)
 - \frac{1}{2} m^2 \beta q \, \delta \Upsilon(\rv, \rv)}
\nonumber\\
&-& n q \rho^0_- \, e^{\beta n  q \Phi(\rv)
 - \frac{1}{2} n^2 \beta q \, \delta \Upsilon(\rv, \rv)}.
\ea
This equation has been derived using field-theoretic methods~\cite{wang,netz-variational,buyukdagli}. It can also be derived using a liquid state theory approach~\cite{bing}.

To see the physical significance of the reaction potential $\Upsilon(\rv,\rv')$, let us consider again inserting a test ion $k q$ at the location $\rv'$ in the electrolyte with Hamiltonian $H_N$.  The Hamiltonian becomes
$$H_N + k q \varphi(\rv'). $$
The {\em conditional} average potential at $\rv$, i.e., the average potential at $\rv$ given the presence of the fixed test ion at $\rv'$ is given by
\ba
\hat{\Phi}(\rv) \equiv \frac{k q }{4 \pi \epsilon |\rv - \rv'|}
   +  \frac{ \Tr_N \, \varphi(\rv)
  \,e^{-\beta  H_N
 - \beta k q \varphi(\rv')}}
 {  \Tr_N   \,e^{-\beta  H_N
 - \beta k q \varphi(\rv')}},
\label{Local-Phi_def}
\ea
where the first term is the direct Coulomb potential due to the test ion, whilst the second term is due to all other ions, whose probability distributions are affected by the presence of the test ion $k q$.  Evidently, in the limit $k q \rightarrow 0$, this potential reduces to the {\em unconditional} average potential $\Phi(\rv)$ (i.e., the average potential at $\rv$ in the absence of any fixed test ion):
\be
\Phi(\rv) = \frac{\Tr_N \, \varphi(\rv) \, e^{-\beta H_N} }
 {\Tr_N \, e^{-\beta H_N} }
= \left \langle \varphi(\rv) \right \rangle_N ,
\label{Phi_def}
\ee
which is precisely what appears in the Poisson equation, Eq.~(\ref{eq:Poisson}), and in the PBE, Eq.~(\ref{eq:pbe}).

If $k q \neq 0$, we can expand Eq.~(\ref{Local-Phi_def}) in terms of $k$.  The first order coefficient then describes the linear response of the average potential at $\rv$ to the insertion of a {\em monovalent} test ion at $\rv'$, which we shall define as the electrostatic {\em Green's function} $\mathcal{G}(\rv, \rv')$:
\be
\hat{\Phi}(\rv)
= \Phi(\rv) + k \, \mathcal{G} (\rv,\rv')  + O(k^2).
\nonumber
\ee
We can calculate $\mathcal{G} (\rv,\rv')$ by taking the derivative of Eq.~(\ref{Local-Phi_def})  with respect to $k$ at $k = 0$:
\ba
\mathcal{G} (\rv,\rv')  &\equiv&
 \left.  \frac{\partial }{\partial k} \hat{\Phi}(\rv)
\right|_{k = 0}
\nonumber\\
&=&   \frac{q }{4 \pi \epsilon |\rv - \rv'|}
- 	\beta  q\, \langle \varphi(\rv)
 \varphi(\rv')\rangle_c
 \nonumber\\
&=&   \frac{q}{4 \pi \epsilon |\rv - \rv'|}
+ \Upsilon(\rv,\rv').
\label{G-Upsilon-relation}
\ea
The physics of the reaction potential $\Upsilon(\rv,\rv')$ now becomes clear: the insertion of the test ion $q$ at $\rv'$ modifies the distribution of all other mobile ions, and hence also changes the potential generated by those ions.  The reaction potential $\Upsilon(\rv,\rv')$ is precisely the part of the electrostatic Green's function that corresponds to this change.

Now, {\em the potential acting on the test ion $k q$ at $\rv$} due to all other ions is given by
\ba
\tilde{\Phi}(\rv) &=& \Limm \left(
      \hat{\Phi}(\rv) -  \frac{k q}{4 \pi \epsilon |\rv - \rv'|}
      \right)
\nonumber\\
&=& \Phi(\rv) + k \,\Upsilon(\rv,\rv).
\label{Phi_til_def}
\ea
The correlation potential $\Upsilon(\rv,\rv)$ is therefore the difference between $\tilde{\Phi}(\rv)$, the local potential acting on a monovalent test ion fixed at $\rv$, and $\Phi(\rv)$, the unconditional average potential at $\rv$.  In other words, the correlation potential of a test ion is the change in the local potential at the position of the test ion that is induced by the test ion's presence.



The FCPBE [Eq.~(\ref{eq:pbe})] is not useful unless we know the correlation potential $\Upsilon(\rv,\rv)$.  There are two possible ways to calculate this quantity.  At a more satisfactory level, we can derive another partial differential equation (PDE) involving both the Green's function $\mathcal{G}(\rv, \rv')$ and the mean potential $\Phi(\rv)$.  This PDE then should be solved self-consistently together with  Eq.~(\ref{eq:mpbe}).  This is usually called {\em the self-consistent Gaussian} approximation, and analytic study of this theory is considerably complicated.  We shall defer study of this theory to a later presentation.  In this work, we shall take a simpler, but cruder approximation, where the average potential is first calculated using the nonlinear PBE (\ref{eq:pbe}), and then the Green's function is calculated using a PBE linearized around the average potential; c.f. Eq.~(\ref{Greens-ODE-0}).  One can then compare these two quantities.  If the correlation potential relative to its bulk value $\delta \Upsilon(\rv,\rv)$ is much smaller than the average potential $\Phi(\rv)$, we can conclude that the former can be ignored in Eq.~(\ref{eq:mpbe}), and therefore the PBE should provide a good approximation.  If, by contrast, the correlation potential  $\delta \Upsilon(\rv,\rv)$ is comparable with, or even larger than the mean potential $\Phi(\rv)$, the PBE then would become qualitatively incorrect, and the self-consistent Gaussian approximation should instead be used.  Our detailed discussion below will make more precise the sense in which $\delta \Upsilon(\rv,\rv)$ can be neglected compared with $\Phi(\rv)$.

The correlation potential of a test ion inside a uniform dilute electrolyte was first calculated by Debye and H\"{u}ckel in their classic work \cite{Debye-Huckel}.  Fixing one ion at the origin, they treated all other ions using {\em the linearized Poisson-Boltzmann theory}, and found that the corresponding correlation potential is given by
\be
\Upsilon_0(\rv,\rv) = - \frac{q}{4 \pi \epsilon \debye}, \label{phi_corr-DH}
\ee
 which is precisely the Coulomb potential generated by an oppositely charged ion at the distance of a Debye length.  Note that the correlation potential is always negative, and moreover, it is linear in the source charge $q$, this being a natural consequence of linearization.  The average Coulomb energy per particle, i.e., the {\em correlation energy}, is then $\varepsilon_{\rm corr} =  q \Upsilon_0(\rv,\rv)/2= - q^2/8 \pi \epsilon \debye$. Proper incorporation of $\varepsilon_{\rm corr}$ into the free energy leads to corrections to the chemical potential and pressure, as well as the equation of state.  These are the essential ingredients of the Debye-H\"{u}ckel theory of electrolytes. For details, see the textbook by Landau and Lifshitz \cite{Landau-SP}.

In this work, we present a generalization of the Debye-H\"{u}ckel method to calculate the correlation potential of a test ion near a \emph{strongly} charged surface inside a dilute electrolyte.  Technical difficulties arise mainly due to the inhomogeneous background potential generated by the charged plate (as well as ions in the bulk).  Analytic results pertaining to the correlation potential for such systems are scarce.  Netz and Orland~\cite{netz_1,netz_2} analyzed the counterion only problem with no discontinuity of permittivity, while Lau~\cite{lau} analyzed the problem of an infinitely thin charged plate inside a $1:-1$ electrolyte.  Both works invoke idealized boundary conditions that ignore image charge effects. The counterion only problem with no dielectric discontinuity has also been studied numerically and in simulations, e.g., by Burak et al.~\cite{burak} On the other hand, using numeric methods, Levin and Flores-Mena~\cite{levin_flores} and Bakhshandeh et al~\cite{bakhshandeh} have analyzed the counterion-only problem for systems with dielectric discontinuity. In this work, we determine the correlation energy for the general case of an $m:-n$ electrolyte, where $m$ and $n$ may or may not be equal, and the dielectric constant of the plate is arbitrary.

The remainder of this paper is organized as follows.  In Sec.~\ref{sec:formalism}, we first define the Green's function and correlation potential, discuss the relevant electrostatic interface conditions, construct the Green's function for a general $m:-n$ electrolyte, and discuss the general properties of the correlation potential in the limit of infinite surface charge density.  In Sec.~\ref{sec:symmetric_electrolyte}, we study the behavior of the correlation energy of the $1:-1$ electrolyte.  In Sec.~\ref{sec:asymmetric} we analyze the corresponding cases of the $2:-1$ and $1:-2$ asymmetric electrolytes.  In Sec.~\ref{sec:general asymmetric} we discuss the general case of an $m:-n$ asymmetric electrolyte.  We finally summarize our results in Sec.~\ref{sec:conclusion}.

\section{Formalism}
\label{sec:formalism}

\subsection{Green's Function}

We follow the original strategy of Debye and H\"{u}ckel, and treat all  ions other than the test ion using linearized PBE.  The important difference is that before the insertion of the test ion, we already have a nonvanishing background potential $\Phi(\rv)$, which must be treated using the nonlinear PBE~(\ref{eq:pbe}).  Upon the insertion of the monovalent test ion at $\rv'$, the average potential is perturbed to $\Phi(\rv) + \mathcal{G}(\rv,\rv^\prime)$, where  the Green's function $\mathcal{G}(\rv,\rv^\prime)$ describes the incremental potential generated by the test ion, together with the resulting reaction of all other ions.  We assume that the perturbation due to the test ion is sufficiently weak, so that the linear response theory is valid.  By taking the first-order variation of Eq.~(\ref{eq:pbe}), we find that the Green's function $\mathcal{G}(\rv,\rv^\prime)$ satisfies the following linearized, inhomogeneous differential equation:
\ba
&-& \epsilon \, \Delta \mathcal{G}(\rv,\rv')
+ \beta q^2 \left( m^2 \rho^0_+  e^{- \beta m q  \Phi }
+  n^2 \rho^0_-  e^{ \beta n  q {\Phi}}
\right) \mathcal{G}(\rv,\rv')
\nonumber\\
&=& q\, \delta(\rv - \rv').
\label{Greens-ODE-0}
\ea
The second term in the left-hand side (LHS) describes the change in distribution of mobile ions, in response to the test ion.

To simplify our notation, let us introduce the following two important length scales:
\begin{subequations}
\ba
\debye &\equiv& \big( \beta q^2
( m^2 \rho^0_+ + n^2 \rho^0_-  )/\epsilon\big)^{-1/2}
\quad \mbox{(Debye length)},
\nonumber\\
\label{Debye_def}\\
b &\equiv&\frac{q^2}{4 \pi \epsilon T}
\quad \quad \quad \quad \quad
\quad \quad  \quad
\mbox{(Bjerrum length)}.
\nonumber\\
\label{Bjerrum_def}
\ea
\end{subequations}
The inverse Debye length is a measure of the strength of screening around the test ion caused by the mobile ions, and the Bjerrum length is the distance between two monovalent ions at which their Coulomb energy equals the thermal energy. Throughout this work, we shall always assume that the electrolyte is sufficiently dilute so that $\ell_{\rm DB}$ is much longer than the Bjerrum length $b$ and  Gouy-Chapman length $\gouy$ [the Gouy-Chapman length will be defined in Eq.~(\ref{mu_def})].
By expressing all lengths in units of $\debye$, and defining the dimensionless potential $\Psi$ as well as the dimensionless Green's function $G(\rv,\rv')$ via
\begin{subequations}
\ba
&& \rv \rightarrow \rv \, \ell_{\rm DB},
\\
&&\Psi \equiv q\beta\Phi,
\label{eq:Psi}
\\
&&G \equiv q\beta\mathcal{G},
\ea
\label{dimensionless_def}
\end{subequations}
Eq.~(\ref{Greens-ODE-0}) can be put in the following much simplified, dimensionless form:
\ba
\bigg[ -  \Delta +
 \frac{m\,  e^{- m \Psi(\rv)}}{m+n}
+\frac{ n\, e^{n \Psi(\rv) }}{m+n}
 \bigg]
 G(\rv,\rv')
= \coupling \, \delta(\rv - \rv'),
\nonumber\\
\label{eq:Green_function_ODE}
\ea
where $\coupling$ is a dimensionless parameter characterizing the importance of the Coulomb energy relative to the thermal energy:
\be
\coupling =  \frac{4 \pi b}{\ell_{\rm DB}}.
\label{coupling-def}
\ee
For a symmetric electrolyte, $m = n $, $\coupling$ is proportional to $\Gamma^{3/2}$, where $\Gamma = (n q)^2/\epsilon T a$ is the Coulomb coupling parameter, and $a$ is the average distance between adjacent ions.  Note that $\coupling$ vanishes in the limit of an infinitely dilute electrolyte, indicating that in this limit, mean field theory (the PBE) becomes exact.

In the bulk electrolyte, $\Phi = 0$, Eq.~(\ref{eq:Green_function_ODE}) reduces to
\be
-  \Delta G_0(\rv,\rv') + G_0(\rv,\rv')
=\coupling \, \delta(\rv - \rv'),
\ee
whose solution is the well-known screened Coulomb (Yukawa) potential:
\be
G_0(\rv, \rv') =  \frac{\coupling\,e^{-|\rv - \rv'|}}{4 \pi |\rv - \rv'|}.
\label{G_0-def}
\ee

\renewcommand{\arraystretch}{1.5}
\begin{center}
\begin{table}[t]
\begin{tabular}{lll }
{\bf Symbol} \quad \quad & {\bf Name}
\quad \quad\quad \quad \quad\quad \quad \quad\quad
\quad \quad\quad   & {\bf Defined in} \\
\hline\hline
$\debye $ & Debye length
& Eq.~(\ref{Debye_def})
\cr
$\gouy$ & Gouy-Chapman length
&Eq.~(\ref{mu_def})
\cr
$\bjerrum$ & Bjerrum length
& Eq.~(\ref{Bjerrum_def})
\cr
$\coupling$ & Dimensionless parameter
& Eq.~(\ref{coupling-def})
\cr
$z_0$ & Location of charged plate
& Fig.~\ref{fig:plate-geometry}
\cr
$\Delta z$ & $=z - z_0$, distance to the plate
& Fig.~\ref{fig:plate-geometry}
\cr
$\Phi(\rv)$ & Dimensionful average potential
& Eq.~(\ref{eq:pbe})
\cr
$w_1(\rv, q)$ & Potential of mean force & Eq.~(\ref{w_1-definition})
\cr
$\mathcal{G}(\rv,\rv')$ & Dimensionful Green's function & Eq.~(\ref{Greens-ODE-0})
\cr
$\Upsilon(\rv,\rv')$ & Dimensionful reaction function & Eq.~(\ref{Upsilon-def})
\cr
$\Upsilon(\rv,\rv)$ & Dimensionful correlation function & Eq.~(\ref{Phi_til_def})
\vspace{2mm}
\cr\hline
$\Psi(\rv)$ & Dimensionless average potential
& Eqs.~(\ref{dimensionless_def})
\cr
$G(\rv,\rv')$ & Dimensionless Green's function
& Eqs.~(\ref{dimensionless_def})
\cr
$G_{0}(\rv,\rv')$ & $G$ in bulk electrolyte & Eq.~(\ref{G_0-def})
\cr
$\chi(\rv,\rv)$ & Dimensionless correlation potential & Eq.~(\ref{chi-rr})
\cr
$\delta \chi(\rv,\rv)$ & $\chi(\rv,\rv)$ relative to its bulk value & Eq.~(\ref{G_G0_relation})
\cr
$\self(\rv)$ & $ = \delta \chi(\rv,\rv)/2$,  Correlation energy
& Eq.~(\ref{corr-energy-def})
\cr
$G(z,z';\kv)$ & F-transformed Green's function
& Eq.~(\ref{eq:fourier_transform_z})
\cr
$G^{\infty}(\rv,\rv^\prime;\kv)$ & $G$ for  infinitely charged plate
& Eq.~(\ref{G_infty_def})
\cr
$\delta \chi^{\infty}(\rv,\rv^\prime;\kv)$ & $\chi$ for  infinitely charged plate
&Eq.~(\ref{self-energy-01})
\cr
$\self^\infty(\rv)$ & $\self(\rv)$  for infinitely charged plate
& Eq.~(\ref{chi-integral})
\cr
\hline
\end{tabular}
\caption{List of frequently-used symbols and their definitions. All quantities in the lower half of the table are dimensionless. }
\label{table:summary}
\end{table}
\end{center}

\subsection{Correlation Potential}
\label{sec:correl}

The correlation potential can be rendered dimensionless by rescaling in units of $T/q$.  Denoting this rescaled correlation potential by $\chi(\rv, \rv)$, we can express it in terms of the Green's function via the following equation [c.f. Eq.~(\ref{G-Upsilon-relation})]:
\ba
\chi(\rv, \rv) &=&
\lim_{\rv' \rightarrow \rv}
\left( G(\rv,\rv')
-  \frac{\coupling}{4 \pi |\rv - \rv'|} \right).
\label{chi-rr}
\ea
Subtraction of the bare Coulomb potential is essential to guarantee the existence of the limit.

The bulk value of the correlation potential can be easily obtained from the bulk Green's function: 
\ba
\lim_{|\rv| \rightarrow \infty}\chi(\rv, \rv)
 &=& \lim_{\rv' \rightarrow \rv}
\left( G_0(\rv,\rv')
-  \frac{\coupling}{4 \pi |\rv - \rv'|} \right)
\nonumber\\
&=&  - \frac{g}{4 \pi}
=  - \frac{\beta q^2}{4 \pi \epsilon \ell_{\rm DB}}
\equiv \chi_0(\rv, \rv).
\label{chi-rr-b}
\ea
It is precisely the dimensionless version of the correlation potential of a monovalent charge in the bulk electrolyte, Eq.~(\ref{phi_corr-DH}).
Subtracting Eq.~(\ref{chi-rr-b}) off from Eq.~(\ref{chi-rr}), we obtain
\ba
\delta \chi(\rv, \rv) &=&
\chi(\rv, \rv) - \lim_{\rv \rightarrow \infty}\chi(\rv, \rv)
\nonumber\\
&=& \lim_{\rv \rightarrow \infty}
\left(
G(\rv, \rv') - G_0(\rv, \rv')
 \right).
\label{G_G0_relation}
 \\
&=& \chi(\rv, \rv) + \frac{g}{4 \pi},
\nonumber
\ea
which is the correlation potential relative to its bulk value.

Finally, let us write the potential of mean force of a $k$-valent ion in its dimensionless form:
\ba
\beta w_1(\rv, k q) &=& k \, \Psi(\rv) + \frac{1}{2} k^2 \Delta \chi(\rv,\rv)
\nonumber\\
 &\equiv& m \, \Psi(\rv) + m^2  \self(\rv).
\ea
The quantity
\be
\self(\rv) =\chi(\rv, \rv) /2
 \label{corr-energy-def}
\ee
is therefore the contribution of ion-ion fluctuation correlations to the potential of mean force of a monovalent ion.  We shall refer to this quantity as the {\em correlation energy}.
Because of the simple relation between $\chi(\rv,\rv)$ and $\self(\rv)$, we shall also use the two terms, correlation potential and correlation energy, interchangeably.

\begin{figure*}
	\centering
		\includegraphics[width=12.5cm]{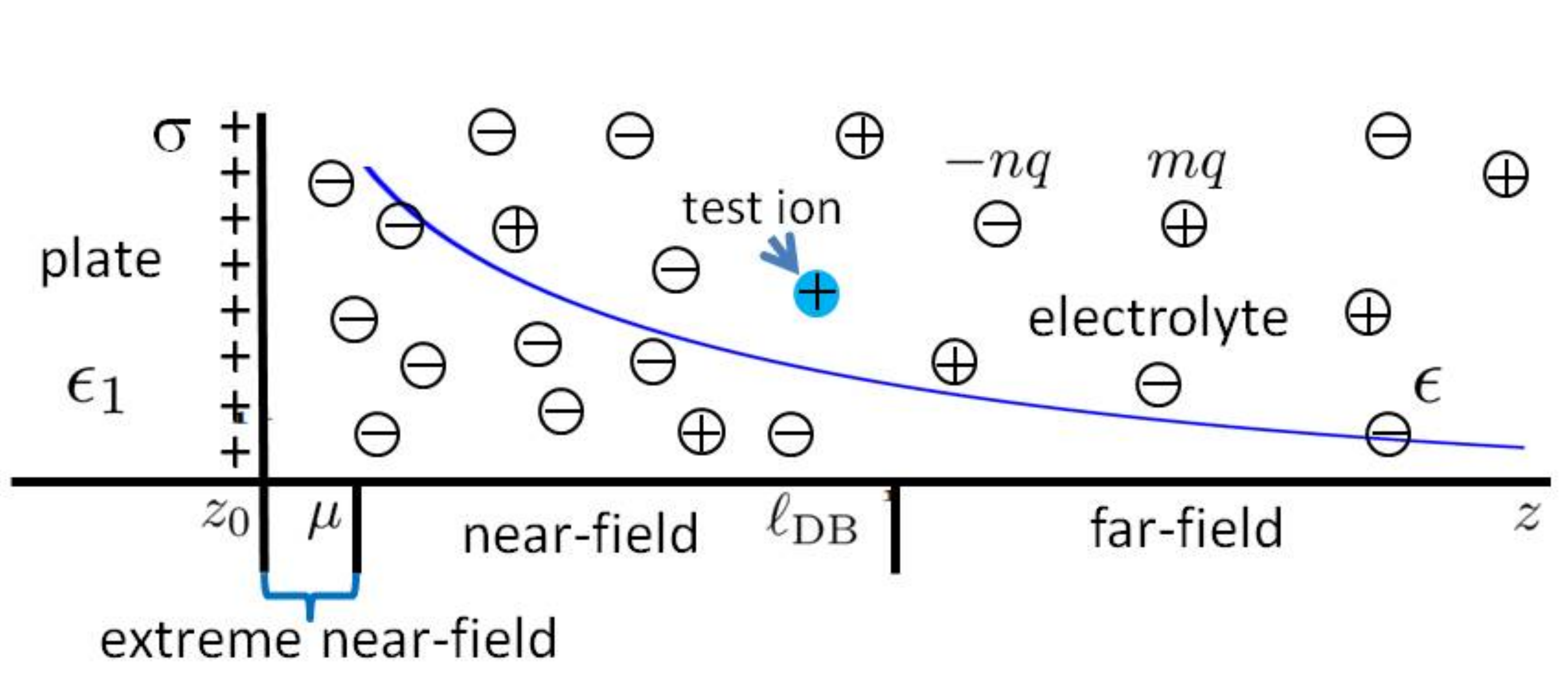}
	\caption{Set-up of the problem: The left half space $z \leq z_0$ is occupied by a dielectric medium with dielectric constant $\epsilon_1$;  the right half space $z > z_0$ is occupied by an $m:-n$ electrolyte with dielectric constant $\epsilon$.  Coions (positive) and counterions (negative) carry charges $+mq$ and $-nq$, respectively. The interface between the dielectric medium and the electrolyte is located at $z=z_0$ and has a uniform positive surface charge density $\sigma$.  The blue-colored ion is our fixed monovalent test ion.  The right half-space $z > z_0$ can be divided into three regions, each characterized by a qualitatively distinct behavior of the correlation energy: (i)~the \emph{extreme near-field region}, typically of the thickness of a Gouy-Chapman length; (ii)~the \emph{near-field region}, typically of the thickness of the order of a Debye length, and (iii)~the \emph{far-field region}, which is the region beyond a Debye length from the interface.  For reference, the mean field potential (scaled in units of $T/q$) for a one-plate system in a $1:-1$ electrolyte is schematically drawn as the blue curve.}
  \label{fig:plate-geometry}
\end{figure*}

\subsection{Interface Conditions}
\label{sec:electrostatic_bc}

The basic geometry of our system is illustrated in Fig.~\ref{fig:plate-geometry}.  A dielectric plate with infinite thickness~\cite{footnote3} is inside an $m:-n$ electrolyte.   $m$ is therefore always the valence of coions in this work.  The dielectric-electrolyte interface is located at $z = z_0$, and carries a uniform {\em positive} surface charge density $\sigma$.  The dielectric constant is $\epsilon_1$ in the left half-space $z < z_0$ and $\epsilon$ in the right half space $z>z_0$.
Although we assume $\sigma > 0$ in this paper, corresponding results can be straightforwardly obtained for a negatively charged plate, by a simple inversion of all charges in the problem.

Note that Eq.~(\ref{eq:pbe}) is the equation satisfied by the mean field potential $\Phi(\rv)$ inside the electrolyte.  Inside the dielectric plate,  $\Phi(\rv)$ satisfies the Poisson equation.   $\Phi(\rv)$ satisfies the free boundary condition at left and right infinities $z = \pm \infty$.  At the interface $z = z_0$, $\Phi(\rv)$ must be continuous, whereas its normal derivative (multiplied by the dielectric constant) has a discontinuity owing to the surface charge density:
\be
- \left. \epsilon_1 \frac{\partial \Phi}{\partial n}\right|_{z_0^-}
+   \left. \epsilon \frac{\partial \Phi}{\partial n}\right|_{z_0^+}
 = \sigma,
\label{BC_dielectric}
\ee
where $z_0^{\pm} = z_0 \pm \varepsilon$, with $\varepsilon$ a positive infinitesimal number.   Note that in the preceding equation, $n$ in the denominators denotes the normal direction to the interface, not the valence of counterions.   We have chosen the unit normal on the interface to point towards the dielectric plate.  The dimensionless version of the interface condition is:
\be
- \left. \epsilon_r \frac{\partial \Psi}{\partial n}\right|_{z_0^-}
+   \left.  \frac{\partial \Psi}{\partial n}\right|_{z_0^+}
 = \eta,
\label{BC_dielectric-dimless}
\ee
where
\be
\epsilon_r = \epsilon_1/ \epsilon
\ee
is the permittivity of the plate relative to that of the electrolyte, and will be referred to as the {\em reduced permittivity}.  $\eta$ is the {\em dimensionless surface charge density}, related to the dimensionful version $\sigma$ via
\be
\eta \equiv \frac{1}{\epsilon} q \beta \sigma \debye
= \frac{2\ell_{\rm DB}}{\gouy},
\label{eta-mu-relation}
\ee
where
\be
\gouy = \frac{2 \epsilon}{\beta q \sigma}
\label{mu_def}
\ee
is the {\em Gouy-Chapman length}, which is a measure of the thickness of the layer of counterions near the plate.  The interface condition for the Green's function $G(\rv,\rv')$ is given by the homogeneous version of Eq.~(\ref{BC_dielectric-dimless}):
\be
- \left.\epsilon_{\rm r} \frac{\partial G}{\partial n}\right|_{z_0^-}
+ \left. \frac{\partial G}{\partial n} \right|_{z_0^+} = 0.
 \label{BC_dielectric_G}
\ee

Because of translational symmetry in the $xy$ plane, the mean field potential $\Phi(\rv)$ depends only on the vertical coordinate $z$.  Furthermore, inside the plate, $\Phi(z)$ depends on $z$ in a linear way.  If we further assume that $\Phi(z)$ is bounded inside the plate, it becomes independent of $z$.~\cite{footnote4}

The location of the interface $z_0$ will be chosen as a function of the surface charge density $\sigma$ such that {\em the mean field potential $\Phi(z)$ is independent of $\sigma$.}   This convention substantially simplifies our analysis, as was demonstrated in Ref.~\cite{xing-ming}.   For large surface charge densities, we can expand $z_0$ as an asymptotic series in powers of $1/\eta$.  For our present purpose, only the leading-order term is needed.  A straightforward analysis (detailed in Sec.~\ref{sec:universality}) shows that
\be
z_0 = \frac{2}{n \, \eta} + O(\eta^{-2}),
\label{z0-eta-general}
\ee
where $n$ is the valence of counterions.  Note that $z_0$ vanishes in the limit of infinite surface charge density.


\subsection{Construction of Green's function}
\label{sec:green_function}

Let us return to the basic geometry illustrated in Fig.~\ref{fig:plate-geometry}. Because of the translational symmetry in the $xy$ plane, the Green's function depends on the transverse coordinates via the combination $\rv_{\perp} - \rv'_{\perp}$.  We can therefore perform a two-dimensional Fourier transform:
\be
\label{eq:fourier_transform_z}
G(\rv,\rv^\prime) = \int \frac{d^2 \kv}{(2 \pi)^2} \,
e^{i\kv \cdot (\rv_{\perp} - \rv_{\perp}^\prime)}
G(z, z';\kv).
\ee
The two-dimensional wave vector $\kv$ is reciprocal to the vector $(\rv_{\perp} - \rv_{\perp}^\prime)$.  Substituting Eq.~(\ref{eq:fourier_transform_z}) into Eq.~(\ref{eq:Green_function_ODE}), we find that $G(z,z';\kv)$ (which we refer to as the F-transformed Green's function, or Green's function for brevity, when there is no danger of confusion) in the electrolyte satisfies the following ordinary differential equation (ODE):
\begin{subequations}
\label{eqs:green}
\ba
\label{eq:green_function_z}
&& \left[ -\frac{d^2}{dz^2} + k^2 +
\left(\frac{ m\, e^{-m\Psi(z)}}{m+n}
+ \frac{n\, e^{n\Psi(z)} }{m+n}\right) \right]
\,
 G(z,z^\prime; \bm{k})
\nonumber
\\
 && \quad \quad \quad \quad\quad \quad \quad\quad
 =  \coupling\, \delta(z-z^\prime)
\quad\quad (z > z_0)
\ea
Inside the plate, $G(z,z';\kv)$ satisfies the Laplace equation, viz.,
\be
\left( -\frac{d^2}{dz^2} + k^2 \right) G(z, z^\prime; k)
= 0
\quad \mbox{ ($z < z_0$)}
\label{eq:green2}
\ee
\end{subequations}
The test ion is always inside the electrolyte, $z'>z_0$.

Equation~(\ref{eq:green2}) has two linearly independent homogeneous solutions $e^{\pm k z}$.  As for Eq.~(\ref{eq:green_function_z}), we first note that $\Psi(z) \rightarrow 0$ in the far-field region; therefore one of the homogeneous solutions to Eq.~(\ref{eq:green_function_z}) must decay as $e^{-\lambda z}$ for large $z$, with $\lambda \equiv \sqrt{1+k^2}$.   We denote this solution by $\phi_-(z)$. The other linearly independent solution then must diverge as $e^{\lambda z}$, and we denote it by $\phi_+(z)$.  Summarizing, we have
\be
\phi_{\pm}(z) \sim e^{\pm \lambda z}, \quad  z \rightarrow \infty.
\label{phi_pm-far}
\ee
Note that $\phi_+(z)$ is determined only up to a linear superposition of $\phi_-(z)$.  Note also that $\phi_{\pm}(z)$ generally depend on the wave vector $\bm{k}$.  For the sake of notational simplicity, however, we do not explicitly display this dependence.

The Green's function $G(z,z';\kv)$ can be constructed using the homogeneous solutions to Eqs.~(\ref{eq:green_function_z}) and (\ref{eq:green2}). In the region $z > z' > z_0$, $G(z, z^\prime; \kv)$ must be proportional to $\phi_-(z)$ in order not to diverge as $z\rightarrow\infty$.  For a similar reason, $G(z, z^\prime; \kv)$ must be proportional to $e^{k z}$ in the region $z < z_0$, in order not to diverge as $z\rightarrow-\infty$.  In the intermediate region ($z_0 < z < z'$), $G(z, z^\prime; \kv)$ is generally a linear combination of the two solutions $\phi_{\pm}(z) $.
 These requirements constrain the functional form of the Green's function to the following:
\ba
&& G(z, z^\prime; \kv)
   \label{eq:form} \\
&=& \left\{ \begin{array}{ll}
 A(z^\prime)\phi_-(z)  &
   \quad \mbox{($z>z^\prime> z_0$)},
   \vspace{3mm}\\
B(z^\prime)\phi_-(z) + C(z^\prime)\phi_+(z) &
   \quad \mbox{( $z^\prime \geq z \geq z_0$)},
   \vspace{3mm}\\
D(z^\prime)e^{k(z-z_0)} &
   \quad \mbox{($z < z_0$)}.
   \end{array}  \right.
   \nonumber
\ea
These three pieces can be patched together using appropriate interface conditions at $z = z_0$ and at $z = z'$.   At $z = z_0$, we have the continuity of $G(z, z^\prime; \kv)$, together with Eq.~(\ref{BC_dielectric_G}):
\begin{subequations}
\ba
&&G(z_0-\varepsilon,z^\prime,\kv)
=G(z_0+\varepsilon,z^\prime,\kv);
\\
&& \epsilon_{\rm r} \frac{d}{dz}G(z_0-\varepsilon,z^\prime,\kv)
=  \frac{d}{dz}G(z_0+\varepsilon,z^\prime,\kv),
\quad\quad
\ea
\label{eq:green_bc_1}
\end{subequations}
where $\varepsilon$ is a positive infinitesimal number.   At $z = z'$, $G(z, z^\prime; \kv)$ is continuous whereas its derivative has a jump as demanded by Eq.~(\ref{eq:green_function_z}):
\begin{subequations}
\ba
&&G(z^\prime-\varepsilon,z^\prime,\kv)
=G(z^\prime+\varepsilon,z^\prime,\kv);
\\
&&\frac{d}{dz}G(z^\prime-\varepsilon,z^\prime,\kv)
-\frac{d}{dz}G(z^\prime+\varepsilon,z^\prime,\kv)
= \coupling.   \quad\quad
\ea
\label{eq:green_bc_2}
\end{subequations}
Solving the four equations~(\ref{eq:green_bc_1}) and (\ref{eq:green_bc_2}) for the four parameters $A$, $B$, $C$, and $D$, we obtain the following expression for the Green's function:
\ba
\label{eq:green_function_general}
G(z, z^\prime; \kv)
&=&  \left\{ \begin{array}{ll}
{\displaystyle
\coupling \, \frac{\phi_-(z^\prime)\, e^{k(z-z_0)}}
{  k \epsilon_r\phi_-(z_0)
-\phi_-^\prime(z_0) }
}
& \quad (z < z_0 < z'),
\vspace{3mm}
\nonumber\\
{\displaystyle
 \coupling \, \frac{\phi_L(z^<)\, \phi_-(z^>)}{W}
 }
&  \quad (z, z^\prime > z_0),
\end{array}\right.
\nonumber\\
\ea
where $z^>, z^<$ are the larger and smaller of $z,z'$; $W$ is the Wronskian of two functions $\phi_{\pm}(z)$, defined by
\be
\label{eq:wronskian}
W \equiv \phi_+(z)\, \phi_-^\prime(z) - \phi_-(z)\, \phi_+^\prime(z).
\ee
The ODE in Eq.~(\ref{eq:green_function_z}) is of Sturm-Liouville type with the second order derivative term having a constant coefficient.  Hence it can be proved that the Wronskian is independent of $z$.~\cite{stone-goldbart} $\phi_L(z)$ is a linear combination of $\phi_\pm(z)$:
\be
\label{eq:phiL_general}
\phi_L(z) \equiv -\phi_+(z) + \phi_-(z)
+ \delta(\kv,z_0,\epsilon_r) \phi_-(z),
\ee
where the dimensionless factor $\delta(\kv,z_0,\epsilon_r)$ is defined as
\be
 \delta(\kv,z_0,\epsilon_r) \equiv
 \frac{k \epsilon_r \phi_+(z_0)
- \phi_+^\prime(z_0)}{k \epsilon_r \phi_-(z_0)
 - \phi_-^\prime(z_0)} - 1.
 \label{delta-factor-def-0}
 \ee
As a comment in passing, we note that even though the function $\phi_+(z)$ is determined only up to a linear superposition of $\phi_-(z)$, the Green's function Eq.~(\ref{eq:green_function_general}) is independent of this arbitrary linear superposition.  We prove this in Appendix \ref{app:gauge-transform}.

Finally, using the F-transformed version of Eq.~(\ref{G_G0_relation}) and Eq.~(\ref{corr-energy-def}), we can express the correlation energy in real space in terms of the following integral over the wave vectors $\kv$:
\ba
\self(z) &=& \frac{{\coupling}}{2}
\int \frac{d^{2}k}{(2\pi)^{2}} \,
\Big( G(z,z;\kv) - G_0 (z,z;\kv)
\Big), \quad
\ea
where
\be
G_0(z,z';\kv) = \frac{\coupling}{2 \lambda} e^{-\lambda|z-z'|}
\ee
is the Fourier transform of $G_0(\rv,\rv')$ [cf. Eq.~(\ref{G_0-def})].

Our task of computing the Green's function and the associated correlation potential is therefore reduced to the calculation of the two homogeneous solutions $\phi_{\pm}(z)$ as well as the associated Wronskian.  We carry out these calculations for different electrolytes separately in Secs.~\ref{sec:symmetric_electrolyte}, \ref{sec:asymmetric}, and \ref{sec:general asymmetric}.

\subsection{Effective Boundary Conditions on the Interface}
\label{sec:BC_revisit}

Using the general expression Eq.~(\ref{eq:green_function_general}) for the Green's function, we can find a relation between its value and its normal derivative on the interface $z = z_0$.  This can be understood as an effective boundary condition for the Green's function.   Let us first consider two limiting cases of $\epsilon_r$, and then consider the general case.

\vspace{2mm}
{\bf The high permittivity limit}, $\epsilon_r \rightarrow \infty$.  We expect that the plate behaves as a conductor.  Indeed, according to Eq.~(\ref{eq:green_function_general}), the F-transformed Green's function inside the plate ($z<z_0$) vanishes in this limit, because the denominator blows up. This is consistent with the fact that the electric field vanishes inside a conductor.  On the other hand, the factor $\delta(\kv,z_0,\epsilon_r)$ in Eq.~(\ref{delta-factor-def-0})  becomes ${\phi_+(z_0)}/{\phi_-(z_0)} - 1$, and hence
$\phi_L(z)$ in Eq.~(\ref{eq:phiL_general}) reduces to
\be
\phi_L(z) \rightarrow  -\phi_+(z) + \phi_-(z) + \left( \frac{\phi_+(z_0)}{\phi_-(z_0)} - 1 \right) \phi_-(z).
\ee
By substituting this into the second line of Eq.~(\ref{eq:green_function_general}) and noting that we are interested in the region $z = z_0^+ < z'$, we find that in the limit $\epsilon_r \rightarrow \infty$,  $G(z ,z'; \kv)$ satisfies the {\em Dirichlet} boundary condition at $z = z_0^+$:
\be
\left. G(z ,z'; \kv)\right|_{z = z_0^+}
 = 0,
  \quad\quad  \epsilon_r \rightarrow \infty.
\label{eq:dirichlet-bc}
\ee
As this ``boundary condition'' holds for all values of $\kv$ and is independent of the wave number $\kv$, it remains valid even if we inverse Fourier transform back to real space.  This confirms our expectation that the potential inside a conductor must be a constant at equilibrium.

\vspace{2mm}
{\bf The low-permittivity limit}, $\epsilon_r \rightarrow 0$. This is a good approximation for most dielectrics inside an aqueous solvent, since typically we have $\epsilon_1 \sim 1, \epsilon \sim 80$.  Equation~(\ref{eq:phiL_general}) in this limit reduces to
\be
\phi_L(z) = -\phi_+(z) + \phi_-(z)
+ \left( \frac{\phi_+^\prime(z_0)}
{\phi_-^\prime(z_0)} - 1 \right) \phi_-(z).
\ee
By substituting this into the second line of Eq.~(\ref{eq:green_function_general}), we find that $G(z ,z'; \kv)$ satisfies the {\em Neumann} boundary condition at $z = z_0^+$:
\be
\label{eq:neumann-bc}
\left. \frac{d}{d z} G^\prime(z,z'; \kv)\right|_{z = z_0^+}
=0,  \quad \quad  \epsilon_r \rightarrow 0.
\ee
Again this condition remains valid even if we inverse Fourier transform back to real space.


\vspace{2mm}
{\bf The general case}, $0 < \epsilon_r < \infty$.  The F-transformed Green's function satisfies the following {\em Robin} boundary condition:
\be
\left.\left( G(z,z'; \kv) - \frac{1}{\epsilon_r k } \frac{d}{dz}G(z,z'; \kv)\right)
\right|_{z = z_0^+}
 = 0.
\ee
It reduces to the Dirichlet boundary condition Eq.~(\ref{eq:dirichlet-bc}) as $\epsilon_r \rightarrow \infty$, and reduces to the Neumann boundary condition Eq.~(\ref{eq:neumann-bc}) as $\epsilon_r \rightarrow 0$.  Note that this effective boundary condition depends explicitly on the wave number $\kv$.  If we inverse Fourier transform back to real space, the resulting Green's function will satisfy a nonlocal effective boundary condition.

\subsection{The Strongly Charged Limit }
\label{sec:strong-limit}
The Green's function exhibits a remarkable property in the strongly charged limit, where
$z_0 \sim \eta^{-1}\rightarrow 0$ [c.f. Eq.~(\ref{z0-eta-general})].  As we show in detail in the following sections, in the strongly charged limit, the two homogeneous solutions to Eq.~(\ref{eq:green_function_z}),
 $\phi_{\pm}(z)$, can be chosen to have the following asymptotic properties as $z_0 \rightarrow 0$:
\begin{subequations}
\label{asymp-phi-phi'}
\ba
\phi_{\pm}(z_0) &=& \frac{1}{z_0} + O(1), \\
\phi'_{\pm}(z_0) &=& - \frac{1}{z_0^2} + O(z_0^{-1}).
\ea
\end{subequations}
Substituting these back into Eq.~(\ref{eq:green_function_general}), and taking the limit  $z_0 \rightarrow 0$ with $z$ fixed, we find that inside the plate $z < z_0$, the Green's function $G(z, z^\prime; \kv)$ scales as $z_0^2$:
\begin{subequations}
\ba
G(z,z';\kv) \sim \coupling \, z_0^2 \,  \phi_-(z^\prime)\,
e^{k(z-z_0)} \rightarrow 0
\nonumber\\
\quad (z_0 \rightarrow 0, \,\,\, z< z_0 \,\,\, \rm{fixed})
\label{G-SCL-1}
\ea
That is, {\em the Green's function vanishes everywhere inside the plate in the limit of infinite surface charge density. }

That the electrostatic potential inside the plate is negligibly small if the surface charge density is very high suggests some profound implications.  Historically, Shklovskii and co-workers \cite{shklovskii2,shklovskii3} have heuristically argued that in the regime of counterion condensation, a strongly charged surface behaves like a conducting surface, because the condensed counterions, being mobile in the lateral directions, form a two-dimensional liquid and are therefore capable of screening out any electrostatic field that might penetrate into the surface. A test ion close to the charged interface therefore should experience an image charge with equal magnitude but opposite sign, which attracts the source ion toward the surface.  This has been argued as the main mechanism driving counterion condensations.  While this argument appears very intuitively convincing, we must be careful when applying it. Near a strongly charged surface, there is indeed a high density of counterions that are mobile in the lateral directions.  These ions however are also mobile along a third direction, perpendicular to the surface.  The way they screen out an external electrostatic field can therefore be very different from that of a two-dimensional ion liquid (emerging in the regime of counterion condensation).  Indeed our analysis of the Green's function below reveals that a test ion near an infinitely charged surface experiences an image charge that is {\em three times bigger} than itself.  This simply cannot happen if the plate behaves as a conductor in the conventional sense.  On the other hand, since our analyses is essentially perturbative in nature, with $\coupling$ treated as a small parameter, it is not clear whether our results apply to the strong-coupling limit.  Detailed analysis using an alternative approach is needed to resolve this issue.

Likewise, because of the asymptotics of Eqs.~(\ref{asymp-phi-phi'}), the factor $\delta(\kv,z_0,\epsilon_r)$ defined in Eq.~(\ref{delta-factor-def-0}) is at least of the order of $z_0$ and vanishes as $z_0  \rightarrow 0$~\cite{footnote5}:
\be
\delta(\kv,z_0,\epsilon_r) = O(z_0).
\ee
 Hence the function $\phi_L(z)$ defined in Eq.~(\ref{eq:phiL_general}) approaches a limiting form:
\be
\lim_{z_0 \rightarrow 0} \phi_L(z) = -\phi_+(z) + \phi_-(z).
\ee
Inside the electrolyte ($z>z_0$), the Green's function [the second line of Eq.~(\ref{eq:green_function_general})] approaches a limiting form:
\ba
\lim_{z_0 \rightarrow 0} G(z,z';\kv)  =
 \frac{\coupling}{W} \big[ -\phi_+(z^<) + \phi_-(z^<) \big] \, \phi_-(z^>)
\nonumber\\
\label{G-SCL-2}
\ea
\end{subequations}
$\phi_{\pm}(z)$ are also independent of $\epsilon_r$, as they are the two homogeneous solutions to Eq.~(\ref{eq:green_function_z}).
It then follows that the {\em Green's function Eq.~(\ref{G-SCL-2}) in the limit of infinite surface charge density is also independent of the permittivity of the plate.  }

For large but finite surface charge density, the correction to the Green's function from the surface charge density is
\ba
\delta G(z,z',\kv) = \frac{\coupling}{W}\delta(\kv, z_0,\epsilon_r) \phi_-(z)\, \phi_-(z').
\label{delta_G-finite}
\ea
The corresponding correction to the correlation energy (relative to the case $z_0 = 0$) can be obtained by equating $z$ with $z'$, and integrating over $\kv$ :
\ba
\self(z) =
\frac{{\coupling}}{2}
\int \frac{d^{2}k}{(2\pi)^{2}} \,
\frac{\delta(\kv,z_0,\epsilon_r) \phi_-(z)^2} {W},
\label{delta_e-finite}
\ea
Even though the factor $\delta(\kv,z_0,\epsilon_r)$ converges to zero as $z_0 \rightarrow 0$, for fixed $\kv$, we shall find that it does not do so uniformly for all wave vectors $\kv$.
Detailed analyses in later sections show that the expansion of the correlation energy in terms of the parameter $z_0$ is a singular one. There is a boundary layer of thickness $z_0$, which we shall call {\em the extreme near-field region}, inside which the perturbation is ill behaved.  The width of this region scales with the Gouy-Chapman length $\gouy$ [recall that $z_0 \sim 1/\eta$ and cf. Eq.~(\ref{eta-mu-relation})], and  shrinks to zero in the limit of infinite surface charge density.  We shall illustrate these properties via explicit calculations for the cases of $1:-1$, $2:-1$, and $1:-2$ electrolytes in Secs.~III D, \ref{sec:asymmetric_positive}, and \ref{sec:asymmetric_negative} respectively, and then analyze the general case of an arbitrary $m:-n$ electrolyte in Sec.~\ref{sec:universality}.

\subsection{Rescaling Transformation}
In its dimensionless form, the nonlinear PBE inside a $m:-n$ electrolyte, Eq.~(\ref{Greens-ODE-0}), is given by:
\be
- \Delta \Psi + \frac{1}{m+n}
\left( e^{n \Psi} - e^{- m \Psi}
\right) = 0.
\label{NLPB-dimless}
\ee
In Ref.~\cite{xing-ming}, it was shown that if $m,n$ have a common factor $p$, such that $m = p\, \tilde{m}, n = p \, \tilde{n}$, then $\widetilde{\Psi} \equiv p\, \Psi$ solves the nonlinear PBE in an $\tilde{m}:\tilde{n}$ electrolyte:
\be
- \Delta \widetilde{\Psi} + \frac{1}{\tilde{m}+\tilde{n}}
\left( e^{\tilde{n} \widetilde{\Psi}} - e^{- \tilde{m} \widetilde{\Psi}}
\right) = 0.
\ee
Note, however, that $\widetilde{\Psi}$ and $\Psi$ satisfy different boundary conditions.  If the surface charge density is $\eta$ for $\Psi$, then it is $p\, \eta$ for $\widetilde{\Psi}$ (assuming, of course, that the charged surface is at the same location for the two cases).

The Green's functions corresponding to these two cases satisfy the equations:
\ba
- \Delta G +  \frac{1}{m+n}
\left( n\, e^{n \Psi} + m \,e^{- m \Psi}
\right) G &=& \coupling  \, \delta(\rv - \rv'),
\nonumber\\
\\
- \Delta \tilde{G} +  \frac{1}{\tilde{m}+\tilde{n}}
\left( \tilde{n}\, e^{\tilde{n} \widetilde{\Psi}}
+ \tilde{m} \,e^{- \tilde{m} \widetilde{\Psi}}
\right) \tilde{G} &=& \coupling \, \delta(\rv - \rv').
\nonumber\\
\ea
Since $m\Psi =  \tilde{m}\widetilde{\Psi}, n\Psi =  \tilde{n}\widetilde{\Psi}$, the preceding two equations are actually identical. Therefore we have the following relation between $G$ and $ \tilde{G}$:
\be
G^{m:-n}(\rv,\rv'; \eta) = G^{\tilde{m}:-\tilde{n}}(\rv,\rv';p\, \eta),
\label{G-tilt_G-relation}
\ee
where $\eta$ and $p\,\eta$ are the dimensionless surface charge densities of the two cases respectively.  As a result, we need to calculate the Green's function only for the cases where $m,n$ are relatively prime.

\section{Symmetric Electrolytes}
\label{sec:symmetric_electrolyte}

In this section we study the correlation potential of a test ion inside a $1:-1$ electrolyte.  Using the relation Eq.~(\ref{G-tilt_G-relation}), we can extend the results to an arbitrary $m:-m$ symmetric electrolyte.  A special version of this problem was previously studied  by Lau~\cite{lau}, where the charged plate is assumed to be infinitely thin, so that image charge effects do not arise.

\subsection{Mean Potential}
Inside a $1:-1$ electrolyte, the PBE~(\ref{NLPB-dimless}) reads
\begin{subequations}
\label{eq:laplace-right}
\ba
&& -  \Psi''(z) + \sinh \Psi(z) = 0, \quad \quad z > z_0, \\
&& - \Psi''(z) = 0, \quad \quad \quad \quad \quad\quad\quad \, z < z_0.
\ea
\end{subequations}
The solution in the right half space $z > z_0$ is well known (see, e.g., \cite{andelman,levin}):
\be
\Psi(z)
= 2 \, \ln \left( \frac{1+ e^{-z}}
{1- e^{-z}}  \right)
= 2 \, \ln \coth \left( \frac{z}{2} \right),
\label{Phi-0}
\ee
The potential in the left half space ($z<z_0$) is a constant.

As in Ref.~\cite{xing-ming}, we choose the value of $z_0$ as a function of the dimensionless surface charge density $\eta$ to fit the interface condition Eq.~(\ref{BC_dielectric-dimless}):
\ba
&&2 \, {\rm{csch}}(z_0) = \eta
\nonumber\\
&&
z_0 = {2}/{\eta} + O(\eta^{-2}).
\label{eq:neumann_2}
\ea
This result of course agrees with the asymptotics of the general case Eq.~(\ref{z0-eta-general}).  Restoring dimensions, we find the following relation between $z_0$ and the Gouy-Chapman length $\gouy$:
\be
\frac{z_0}{\debye} = \ln\left(
 \sqrt{1+ \frac{\gouy^2}{\debye^2}}
+\frac{\gouy}{\debye} \right).
\ee
For high surface charge density, we have  $z_0 \approx \gouy \ll \ell_{\rm DB}$.

\subsection{Green's function}
We now proceed to evaluate the Green's function.  Setting $m = n = 1$,  Eqs.~(\ref{eqs:green}) reduce to the following forms:
\begin{subequations}
\ba
 \left( -\frac{d^2}{dz^2} + \lambda^2 +
\frac{2}{\sinh^{2}z}\right) G(z, z^\prime; \kv)
 &=& \coupling \, \delta(z-z^\prime)
 \nonumber\\
  \mbox{ ($z > z_0$)}, &&
\label{eq:green1} \\
\vspace{3mm}
 \left( -\frac{d^2}{dz^2} + k^2 \right) G(z, z^\prime; \kv) &=& 0
 \nonumber\\
\mbox{ ($z < z_0$)},  &&
\ea
\end{subequations}
where $\lambda \equiv \sqrt{1+k^2}$.
Equation~(\ref{eq:green1}) has two independent homogeneous solutions
 \begin{subequations}
\label{eq:phi+phi-1:-1}
\ba
\phi_+(z) &=& (\coth(z) - \lambda) e^{\lambda z},
\\
\phi_-(z) &=& (\coth(z) + \lambda) e^{-\lambda z}.
\ea
\end{subequations}
These solutions exhibit the far-field asymptotics Eq.~(\ref{phi_pm-far}), as well as the near-field asymptotics Eqs.~(\ref{asymp-phi-phi'}), as we demanded earlier.  The Wronskian formed by $\phi_{\pm}(z)$ is independent of $z$:
\be
W = \phi_+(z)\, \phi_-^\prime(z) - \phi_-(z)\, \phi_+^\prime(z)
= 2\lambda(\lambda^2-1).
\label{W:1:-1}
\ee

To obtain the F-transformed Green's function, we substitute Eqs.~(\ref{eq:phi+phi-1:-1}), (\ref{W:1:-1}) into Eqs.~(\ref{delta-factor-def-0}), (\ref{eq:phiL_general}), and (\ref{eq:green_function_general}).   For the field point inside the plate $z<z_0$, we have:
\begin{subequations}
\ba
&&G(z, z^\prime; \kv)
= 2\lambda(\lambda^2-1) \times
\nonumber\\
&&\frac{ \coupling\,  e^{k (z- z_0) + \lambda z_0 } \phi_-(z^\prime)}
{\left(\coth (z_0) (k \epsilon_r +\lambda)+\lambda (k \epsilon_r + \lambda)
+{\rm{csch}}^2(z_0)\right)} \nonumber \\
&&\quad\quad\quad\quad\quad\quad\quad\quad\quad\quad
\quad\quad (z<z_0).
 \label{G-in-plate-1:-1}
\ea
In the limit of infinite surface charge density, the Green's function Eq.~(\ref{G-in-plate-1:-1}) vanishes as $z_0 \rightarrow 0$, because the denominator blows up as $z_0^{-2}$.

For the field point inside the electrolyte $z > z_0$, we have:
\ba
G(z, z^\prime; \kv)
&=& \frac{\coupling}{2\lambda(\lambda^2-1)}
 \phi_-(z^>) \Big( - \phi_+(z^<) + \phi_-(z^<)
\nonumber\\
&+&  \delta(\kv,z_0, \epsilon_r) \phi_-(z^<) \Big),
 \quad\quad
(z > z_0),
 \label{eq:Gl_rewrite}
\ea
\end{subequations}
where $z^<$ and $z^>$ are the larger and smaller of $z$ and $z^\prime$.  The factor $\delta(\kv,z_0,\epsilon_r)$ [defined in Eq.~(\ref{delta-factor-def-0})] is
\be
\delta(\kv,z_0,\epsilon_r) =
 \frac{(\lambda - k \epsilon_r)(\lambda - \coth(z_0))
+ {\rm{csch}}(z_0)^2}{(\lambda + k \epsilon_r)
(\lambda + \coth(z_0))+{\rm{csch}}(z_0)^2}
e^{2 \lambda z_0} - 1.
\label{delta-factor-def}
\ee
Note that the test ion is always in the electrolyte $z^\prime > z_0$.

\subsection{Infinite Surface Charge Density}
\label{subsec:sym_infinite}

For the case of infinite surface charge density, we substitute Eqs.~(\ref{eq:phi+phi-1:-1}) into (\ref{eq:Gl_rewrite}), take the limit $z_0 \rightarrow 0$, and further equate $z' = z$:
\ba
&&  G^\infty(z, z; \kv)
\equiv
\lim_{\eta \rightarrow \infty} G(z, z; \kv)
\label{G_infty_def}
\label{self-energy-0}
\\
&=&  \frac{ \coupling
 \left(
(\lambda + \coth z)^2 e^{- 2 \lambda z}
+ ( \lambda^2 - \coth^2 z )
\right)
}{2 \lambda (\lambda^2 - 1)},
\quad (z > 0).
\nonumber
\ea
Here, the superscript \lq\lq$\infty$\rq\rq\ refers to the fact that the plate is infinitely-charged.

Now, as $z \rightarrow \infty$, the Green's function approaches the value
\be
\lim_{z \rightarrow \infty} G^\infty(z,z;\kv)
= \frac{\coupling}{2 \lambda}
= G_0(z,z;\kv).
\label{G0-FT}
\ee
But this is exactly the F-transformed Green's function in the bulk electrolyte, which can be obtained from Eq.~(\ref{G_0-def}).  This result, of course, applies to arbitrary types of electrolyte.

Subtracting Eq.~(\ref{G0-FT}) from Eq.~(\ref{self-energy-0}), we obtain
\ba
&& \delta \chi^\infty(z,z; \kv) = G^\infty(z,z;\kv)
-  G_0(z,z;\kv)
 \label{self-energy-01}
\\
&=&  \frac{\coupling}{2\, \lambda (\lambda^2 - 1)}
   \left[  { (\coth (z)+\lambda )^2 e^{-2 z \lambda }
   -  \text{csch}^2(z) } \right].
 \nonumber
\ea
This is the Fourier space version of  Eq.~(\ref{G_G0_relation}) [c.f. also Eq.~(\ref{eq:fourier_transform_z})].  To obtain the correlation energy in real space, we have to integrate Eq.~(\ref{self-energy-01}) over $\kv$:
\be
\self^\infty(z) =
\frac{1}{2}
\int \frac{d^2 \kv}{(2 \pi)^2}\chi^\infty(z,z;\kv).
\label{chi-integral}
\ee
The integral over $\kv$ turns out to be quite subtle, but the final result is rather simple.  We relegate the details of the calculation to Appendix \ref{app:integral:1:-1} and exhibit the result directly:
\ba
&&  \self^\infty(z)  =
\label{self-energy-1} \\
&& \frac{\coupling} {8 \pi}
 \Bigg[
\frac{e^{-2 z}}{2 z}
- \frac{1}{2} \text{csch}^2(z)
 \big( \ln (4z) + E_1(4 z)+\gamma \big)
\Bigg],\nonumber
\ea
where
\be
 E_1(z) \equiv \int_1 ^{\infty} t^{-1} e^{-t z} dt
= \int_z^{\infty} u^{-1} e^{-u }du
\ee
is a {\em generalized exponential integral function}, and $\gamma = 0.5772\cdots$ is the Euler-Mascheroni constant.  This correlation energy is negative for all values of $z$, increasing monotonically towards zero as $z \rightarrow \infty$.

\begin{figure}
	\centering
		\includegraphics[width=8.5cm]{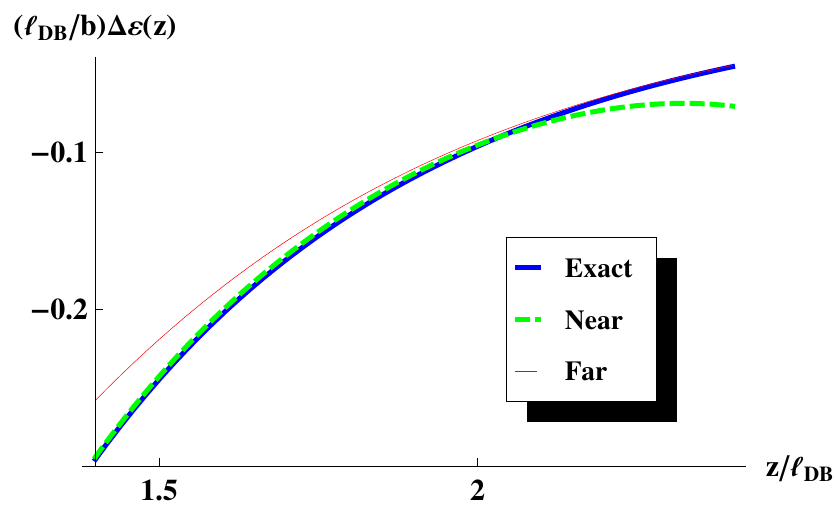}
	\caption{(Color online) {\bf 1:-1 electrolyte} \,\, The correlation energy of an ion (scaled in units of $b/\debye$) near an infinitely charged plate, for the region $1.4 < z < 2.5$.   The exact result is shown as the blue thick solid curve, the near-field expansion Eq.~(\ref{selfenergy-nearfield}) up to the order of $z^6$ is shown as the green, dot-dashed curve, and the leading-order far-field expansion Eq.~(\ref{selfenergy-farfield}) is shown as the red thin solid curve.  Both approximations match well with the exact result around $z \approx 2$.  Correction due to the finiteness of surface charge density is negligible in this region, as long as $z_0 \ll 1$.}
  \label{fig:selfenergy1-1}
\end{figure}

\vspace{2mm}
\textbf{The near-field region}.
Let us look at the near-field asymptotic expansion of the correlation energy Eq.~(\ref{self-energy-1}).  Up to the order of $z^6$, we have
\ba
  \self^\infty(z)
  &=& \frac{\coupling}{8 \pi}
  \Big(
 - \frac{3}{2 z}+1-\frac{z}{9}-\frac{41 z^3}{675}
 +\frac{4 z^4}{135}
 \nonumber\\
&-&  \frac{22 z^5}{33075}
 -\frac{4 z^6}{2835}
 +O\left(z^7\right)
 \Big).
\label{selfenergy-nearfield}
\ea
The first seven terms  provide a remarkably accurate approximation to the self-energy for the whole range of $0 < z < 2 \, \ell_{\rm DB}$, as shown in Fig.~\ref{fig:selfenergy1-1}.
Also shown in this figure is the leading-order far-field expansion (red thin solid line) and the exact result (blue thick solid line). 

The leading term of the above near-field expansion
\be
-\frac{3 g}{2 \cdot 4 \pi (2z)} = -\frac{3b}{\debye(4z)}
\ee
can be interpreted as arising from an \lq\lq image charge\rq\rq\ of magnitude $-3q$ located at a distance $z$ behind the plate.  We must emphasize that this \lq\lq image charge\rq\rq\ is not a consequence of discontinuity in permittivity, as in the usual electrostatic interface problems, because the reduced permitivity $\epsilon_r$ does not even show up in our result.  Rather, the ``image charge'' emerges from the screening effects of counterions accumulated near the strongly charged surface.  The fact that the ``image charge'' is three times bigger than the test charge is rather intriguing, but clearly shows that a strongly charged interface is essentially different from a conventional conductor surface.  We shall explore its implications in depth in a separate presentation.

\vspace{2mm}
\textbf{The far-field region}.
The far-field expansion of the correlation energy is also interesting.  To the leading-order
we have
\ba
&&  \self^\infty(z) =
 \label{selfenergy-farfield} \\
&&
\frac{\coupling} {8 \pi} \bigg(
\left( - 2 \gamma - 2 \ln (4z)
+ \frac{1}{2 z} \right) e^{-2 z}
+  O\left( e^{-3z}\right) \bigg).
\nonumber
\ea
 As shown in Fig.~\ref{fig:selfenergy1-1}, this leading-order approximation is excellent for $z >2$.  In the far-field region, the correlation energy, i.e., equivalently, the interaction energy between a test ion and a charged surface, is doubly screened, decaying as $e^{-2z/\ell_{\rm DB}}$ (restoring dimensions), and therefore is much smaller than the mean field electrostatic potential energy, which scales  as $e^{-z/\ell_{\rm DB}}$.   Consequently, inside a symmetric electrolyte, PB theory should constitute a good approximation in the far-field region.

\begin{figure}[t]
	\centering
		\includegraphics[width=8.5cm]{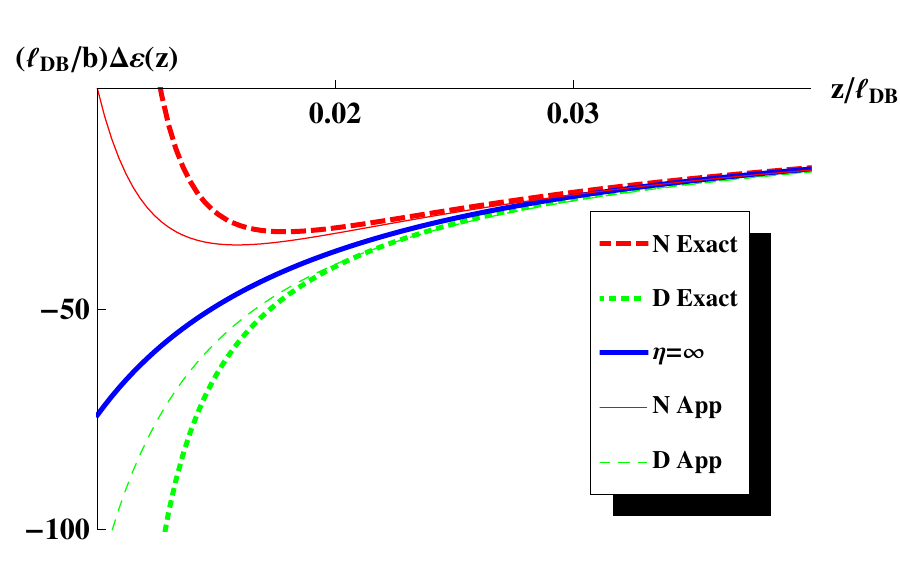}
	\caption{ (Color online) {\bf Near- and Extreme Near-Field }\,\,
	The correlation energy of a test ion in the near-field and the extreme near-field regions.
Valences of counter- and coions play no role in these regions.  Except for the curve ``$\eta = \infty$'', we have chosen $z_0 = 0.01\debye$, corresponding to a dimensionless surface charge density $\eta = 200$, and the plate is located at $z = z_0$.
\lq\lq N Exact\rq\rq and \lq\lq D Exact\rq\rq show direct numerical integration of Eq.~(\ref{eq:Gl_rewrite}) with $\epsilon_r = 0$ and $\infty$, respectively.
Note that these two curves diverge towards $+\infty$ and $-\infty$, respectively, due to conventional image charge effects; cf. Eq.~(\ref{chi-extreme-near}).
\lq\lq $\eta=\infty$\rq\rq represents the correlation energy with an infinitely charged plate at the origin; cf. Eq.~(\ref{self-energy-1}).
This correlation energy remains finite at $z = 0$.
\lq\lq N App\rq\rq and \lq\lq D App\rq\rq show the sum of the near-field approximation~(\ref{selfenergy-nearfield}) and the leading-order (in $z_0$) correction~(\ref{finite-near-field}), with $\theta = 1$ and $-1/2$, respectively.  Note that these correlation energies also remain finite as $z \rightarrow z_0$.  Perturbation in terms of $z_0$ fails in the extreme near-field $0< z - z_0 < z_0$.
Note also that all curves converge to a single curve as $z \gg z_0$, demonstrating that boundary conditions (i. e., permittivity of the plate) plays no role except in the extreme near-field.}
  \label{fig:selfenergy1-2}
\end{figure}

\subsection{Finite Surface Charge Density}
\label{sec:1:1_finite}
If the surface charge density is large but finite, the factor $\delta(\kv,z_0,\epsilon_r)$ does not vanish.  To obtain the correction to the correlation potential (relative to the case $z_0 = 0$), we would have to calculate the integral Eq.~(\ref{delta_e-finite}), where various ingredients in the integrand are given by Eqs.~(\ref{eq:phi+phi-1:-1}), (\ref{W:1:-1}), and (\ref{delta-factor-def}), respectively.
Unfortunately, we are not able to calculate this integral in a closed form.  We shall therefore expand $\delta(\kv,z_0,\epsilon_r)$ defined in Eq.~(\ref{delta-factor-def}) in terms of the small parameter $z_0$, and then carry out the integral  Eq.~(\ref{delta_e-finite}) term by term.



 The expansion in terms of $z_0$, however, depends on the value of the reduced permittivity $\epsilon_r$.  For $\epsilon_r \ll 1/z_0$, we directly expand Eq.~(\ref{delta-factor-def}) in terms of $z_0$:
\be
\delta(\kv,z_0,\epsilon_r) =
\lambda(\lambda^2-1) z_0^3
\Big( \frac{4}{3}
 - 2 \epsilon_r k z_0 \Big)
 + O(z_0^5).
\ee
For $\epsilon_r \gg 1/z_0$, by contrast, we should first take the limit $\epsilon_r \rightarrow \infty$, and then expand in terms of $z_0$:
\be
\delta(\kv,z_0,\epsilon_r) = - \frac{2}{3}
\lambda(\lambda^2-1) z_0^3 + O(z_0^5).
\label{delta-z0-expansion}
\ee
Substituting these back into Eq.~(\ref{delta_e-finite}), we find that, to the order of $z_0^3$, the correction to the correlation potential is given by
\ba
\label{eq:green_correction}
\self(z)  &=&
\frac{1}{2} \int \frac{d^{2}k}{(2\pi)^{2}}
\left(  \frac{2 }{3}  \theta \coupling z_0^3
 + O(z_0^4) \right)
 \nonumber\\
& \times& (\lambda + \coth(z))^2 e^{-2\lambda z}
\nonumber\\
&\approx&  \frac{\theta \, \coupling}{192\pi}
 \cdot \frac{z_0^3} {z^4}  \text{csch}^2(z)
 \Big(
(2 z+1) \left(8 z^2+4 z + 3 \right)
\nonumber\\
&-& 2 e^{-2 z} \left(4 z^2+6 z+3\right)
+ e^{-4 z} (2 z+3)  \Big).
\label{eq:correction_to_correlation_potential}
\ea
where $\theta = 1$ for $\epsilon_r z_0 \ll 1$ (insulator plate) and $\theta = -1/2$ for $\epsilon_r z_0 \gg 1$ (conductor plate).


The far-field expansion of Eq.~(\ref{eq:correction_to_correlation_potential}) reads
\ba
\delta \varepsilon(z) &=&
\frac{\theta\,\coupling z_0^3 }{48\pi}
 \left[ \left( \frac{16}{z} + \cdots
\right) e^{-2z}
 +  O(e^{-3z}) \right],
 \quad\quad
\label{delta_e-finite_far}
\ea
where the ellipsis refers to terms of the order $z^{-2}$ and lower.  $\delta \varepsilon(z)$ is smaller than the leading-order result Eq.~(\ref{selfenergy-farfield}) by a factor of $z_0^3$, and therefore is negligible in the strongly charged regime.  In the near-field region $z\ll1$, Eq.~(\ref{eq:correction_to_correlation_potential}) can be expanded in terms of small $z$:
\ba
\self(z) =
\theta \cdot \frac{3\coupling}{16\pi}\cdot
\left( \frac{ z_0}{z} \right)^3 \left( \frac{1}{z}
+ O(1) \right).
\label{finite-near-field}
\ea
This correction is smaller than the leading-order result Eq.~(\ref{selfenergy-farfield}) by a factor of $(z_0/z)^3$, and therefore is also negligible as long as $z \gg z_0$.  However, if the field point is very close to the plate, $z \approx z_0$, the correction Eq.~(\ref{finite-near-field}) scales as $ \coupling/z_0$ and therefore is of the same order as the leading-order result in Eq.~(\ref{selfenergy-nearfield}).  This suggests a break-down of perturbation theory in powers of $z_0$.  Indeed, we can work out the higher-order terms  in our expansion Eq.~(\ref{delta-z0-expansion}) in terms of $z_0$.  The resulting higher order corrections $\self(z)$ all scale as $\coupling/z_0$  for $z \approx z_0$.  The expansion in terms of $z_0$ does not converge at $z = z_0$.  Our perturbation in terms of $z_0$ is therefore a {\em singular} one.  In Fig.~\ref{fig:selfenergy1-2}, we compare the exact correlation energy [via numerical integration of Eq.~(\ref{eq:Gl_rewrite})] with the sum of Eq.~(\ref{selfenergy-nearfield}) and Eq.~(\ref{finite-near-field}).  It is clear from this figure that perturbation in terms of $z_0$ breaks down in the extreme near-field region.

\vspace{2mm}
\textbf{The extreme near-field region} ($0 < z -z_0 \ll z_0$). The above analysis shows that there is an \emph{extreme near-field} region where $z$ is of comparable magnitude to $z_0$, and the perturbation in terms of $z_0$ breaks down.  To obtain the asymptotics of the correlation energy in this region, we need to perform a different analysis.  The details are relegated to Sec.~\ref{sec:universality}.
Here, we simply state the result, viz.,
\be
\self(z)
\approx
\frac{\coupling}{4\pi}
\cdot
\frac{1 - \epsilon_r }{1 + \epsilon_r}
\cdot
\frac{1}{4(z-z_0)}.
 \label{chi-extreme-near}
\ee
This is precisely the interaction energy between the test ion and a neutral dielectric interface with relative permittivity $\epsilon_r$, as can be found in  standard textbooks on electrostatics \cite{jackson,griffiths}.  As is well known, this interaction can be interpreted as arising from an image charge $q(1-\epsilon_r)/(1+\epsilon_r)$ at the symmetric point.  The distance between the ion and the interface is $z-z_0$, whereas that between the test ion and the image charge is $2(z-z_0)$.  Therefore in the extreme near-field region, the correlation energy of the test ion is dominated by the discontinuity in permittivity, with all other ions playing a less important role.  Since $\epsilon_r >0$ for all normal dielectrics, the magnitude of this image charge is always less than that of the source ion.  In Sec.~\ref{sec:universality}, we show that the extreme near-field asymptotics Eq.~(\ref{chi-extreme-near}) actually holds  for arbitrary valences $m:-n$.

Is this extreme near-field region relevant to real systems? To answer this question, we must remember that in reality ions are not point-like. Instead they have some finite hard core radius $a$, which sets a minimal distance between them and a charged interface. This radius is typically a few angstroms inside an aqueous solvent. The extreme near-field region is accessible only if the Gouy-Chapman length $\gouy$ is longer than the ion radius.

We now summarize the behaviors of the correlation energy of a test ion inside a \emph{symmetric} electrolyte in three different regions: (i)~In the \emph{far-field} region ($z\gg \ell_{\rm DB}$, restoring dimensions), the correlation energy [Eq.~(\ref{selfenergy-farfield})] is doubly screened. (ii)~In the \emph{near-field} region (but not too close to the plate, $\ell_{\rm DB} \gg z - z_0 \gg \gouy $), the correlation energy [Eq.~(\ref{selfenergy-nearfield})] can be interpreted (in the limit of infinite surface charge density) as the interaction energy between the source ion and a point image charge of strength $-3 q$.  (iii)~In the \emph{extreme near-field} region ($z - z_0 \ll z_0 \sim \gouy$), the correlation energy is dominated by discontinuity of the permittivity [cf. Eq.~(\ref{chi-extreme-near})].  (iv)~The correction due to the finiteness of surface charge density is negligible, except in the extreme near-field region.  We shall see below that results~(ii), (iii), and (iv) also hold for an \emph{asymmetric} electrolyte, whereas result~(i) is essentially modified.

\section{Asymmetric Electrolytes: $2:-1$ and $1:-2$}
\label{sec:asymmetric}
The analyses for the cases of $2:-1$ and $1:-2$ asymmetric electrolytes are analogous to that of the symmetric electrolyte, but are technically much more involved.  We shall discover that in these \emph{asymmetric} electrolytes, the correlation energy decays as $e^{-z}$ in the far-field, that is, it is \emph{singly} screened.  The significance of this result will be discussed in Sec.~\ref{sec:failure-PB}.

\subsection{Mean Potential}
The PBEs for the $2:-1$ and $1:-2$ asymmetric electrolytes are given (in dimensionless form) respectively by
\begin{subequations}
\label{PBeqns-2:1}
\ba
-\Delta \Psi + \frac{1}{3} \left(
e^{\Psi} - e^{-2 \Psi}
\right) &=& 0, \quad (2:-1);\\
-\Delta \Psi + \frac{1}{3} \left(
e^{2 \Psi} - e^{- \Psi}
\right) &=& 0, \quad (1:-2).
\ea
\end{subequations}
The potentials generated by an isolated positively charged plate are, respectively:
\begin{subequations}
\label{PBsolution-2:1}
\ba
\Psi^{2:-1}(z) &=& \ln  \frac{1+ 4 \, e^{-z} + e^{-2z}}
{\left(1- e^{-z} \right)^2} ;
\label{PBsolution-2:1+}
\\
\Psi^{1:-2}(z) &=& \ln  \frac
{ \left( 1 +  e^{-z_1 } \right)^2 }
{ 1- 4 \, e^{-z_1 }
+ e^{-2z_1 } },
\label{PBsolution-2:1-}
\ea
\end{subequations}
where $z _1 = z + \ln(2+ \sqrt{3})$.  Both solutions exhibit a logarithmic singularity at $z = 0$.  $\Psi^{1:-2}(z)$ differs from the result in Ref.~\cite{xing-ming} by a trivial translation of $z$.

As in the $1:-1$ case, a finitely charged plate is located at $z_0$, which is chosen such that the potentials Eq.~(\ref{PBsolution-2:1}) become independent of $z_0$.  This determines $z_0$ as a function of surface charge density $\eta$ via
\be
\frac{\partial \Psi}{\partial z}(z_0) =  - \eta.
\label{Neumann-1}
\ee
Using  Eqs.~(\ref{Neumann-1}) and (\ref{PBsolution-2:1}), we find that to the leading-order
\ba
z_0  &=& {2}/{\eta} + O(\eta^{-2}), \quad (2:-1);\\
z_0 &=& {1}/{\eta} + O(\eta^{-2}), \quad (1:-2),
\ea
which agree with the general result Eq.~(\ref{z0-eta-general}).


\subsection{$2:-1$ electrolyte}
\label{sec:asymmetric_positive}

In a $2:-1$ electrolyte ($z > z_0$), the Green's function satisfies the linearized inhomogeneous PBE, whilst inside the plate ($z < z_0$), it obeys the Laplace equation:
\begin{subequations}
\ba
&&- \frac{d^2}{dz^2} G(z,z';\kv) + \left[
k^2 +
\frac{1}{3} \left(
e^{\Psi} + 2 \, e^{-2 \,\Psi}
\right)
\right]G(z,z';\kv)
\nonumber \\
&& \quad \quad \quad \quad \quad \quad
\quad \quad \quad \quad
 = \coupling
\, \delta(z-z') \quad \quad\quad\mbox{($z > z_0$)},
\nonumber\\
\label{Green-equation-2:1}
\\
&& \left( -\frac{d^2}{dz^2}
+ k^2 \right) G(z, z^\prime; \kv)
= 0
\quad\quad\quad\quad \quad \,\,\,\, \mbox{ ($z < z_0$)},
\nonumber\\
\ea
\end{subequations}
where the mean field potential $\Psi(z)$ is given by Eq.~(\ref{PBsolution-2:1+}).
As before, in order to obtain the Green's function, we first need to find two independent homogeneous solutions $\phi_+(z)$ and $\phi_-(z)$ to Eq.~(\ref{Green-equation-2:1}). It is remarkable enough that these solutions can be expressed in terms of elementary functions:
\begin{subequations}
\label{eq:phi2:1positive}
\ba
 \phi_+ (z) &=& - \frac{1}{2 \lambda} e^{\lambda z }
\Bigg[
{(\lambda -1)(2 \lambda -1)}
\\
&+&
\frac{6\,e^{-z} (\lambda - 1
- (2\lambda - 1) e^{-z} - \lambda e^{-2z}  )}
{(1-e^{-z})(1+ 4\,e^{-z} + e^{-2z})}
\Bigg];
\nonumber\\
 \phi_- (z) &=& \frac{1}{2 \lambda} e^{-\lambda z }
\,\Bigg[
{(\lambda +1)(2 \lambda +1)}
\\
&-&
\frac{6\,e^{-z} (\lambda + 1
- (2\lambda + 1) e^{-z} - \lambda e^{-2z}  )}
{(1-e^{-z})(1+ 4\,e^{-z} + e^{-2z})}
\Bigg].
\nonumber
\ea
\end{subequations}
It is easy to check that these solutions exhibit the near-field asymptotics Eqs.~(\ref{asymp-phi-phi'}) as well as the far-field asymptotics Eqs.~(\ref{phi_pm-far}), as we demanded earlier.   The Wronskian formed by $\phi_{\pm}$  can be easily calculated using Eq.~(\ref{eq:wronskian}):
\ba
W(\phi_+, \phi_-)
=  \frac{1}{2\lambda}
(4\lambda^4-5\lambda^2+1).
\label{eq:wronskian_positive_plate}
\ea
To obtain the F-transformed Green's function, we substitute Eqs.~(\ref{eq:phi2:1positive}) and (\ref{eq:wronskian_positive_plate}) into Eqs.~(\ref{delta-factor-def-0}), (\ref{eq:phiL_general}), and (\ref{eq:green_function_general}).
 We shall however not write it out in detail as it is rather bulky and complicated.

\subsubsection{Infinite Surface Charge Density}

For an infinitely charged surface, $z_0 = 0$, and the F-transformed Green's function is given by Eq.~(\ref{G-SCL-2}), with $\phi_{\pm}$ given by  Eqs.~(\ref{eq:phi2:1positive}).  Subtracting the Green's function in the bulk, Eq.~(\ref{G0-FT}), we find the F-transformed correlation potential as
\ba
\delta \chi^\infty(z,z;\kv) &=&
 G^\infty(z,z; \kv) - G_0(z,z; \kv)
\label{selfenergy-k-2:1} \\
&=&
\frac{\coupling\left(
 - \phi_+(z)  \phi_-(z) + \phi_-(z)^2 \right)}
 {(4 \lambda^4 - 5 \lambda^2 +1)/(2 \lambda) }
 - \frac{\coupling}{2 \lambda},
\nonumber
\ea
with $\phi_{\pm}(z)$ defined in Eqs.~(\ref{eq:phi2:1positive}).   By integrating over the wave vector $\kv$, we obtain the correlation energy $\self^\infty(z)$ for a monovalent test ion positioned at $z$.  The (very complicated) full expression, together with details of the calculation, is displayed in Appendix~\ref{app:integral:2:1_positive_plate}.   Here, we present its near-field and far-field asymptotic series.

\vspace{2mm}
\textbf{The near-field region}. The near-field expansion of $\self^\infty(z)$ is given by
\be
\label{eq:bare_self_energy_positive_plate}
\self^\infty(z)
 =
 \frac{g}{8 \pi}
 \left( -\frac{3}{2z}+ 1- \frac{z}{18}
 + O(z^2) \right).
\ee
The first two terms of this series are identical to those for the $1:-1$ electrolyte in Eq.~(\ref{selfenergy-nearfield}).  In fact, in the region plotted in Fig.~\ref{fig:selfenergy1-2}, Eq.~(\ref{eq:bare_self_energy_positive_plate}) is virtually indistinguishable from the corresponding result for the $1:-1$ electrolyte, Eq.~(\ref{selfenergy-nearfield}).  In Sec.~\ref{sec:universality}, we show that for an infinitely charged plate, the leading-order near-field asymptotics of the correlation energy is independent of the valences of counterions and coions.

\begin{figure}[t]
	\centering
		\includegraphics[width=8.5cm]{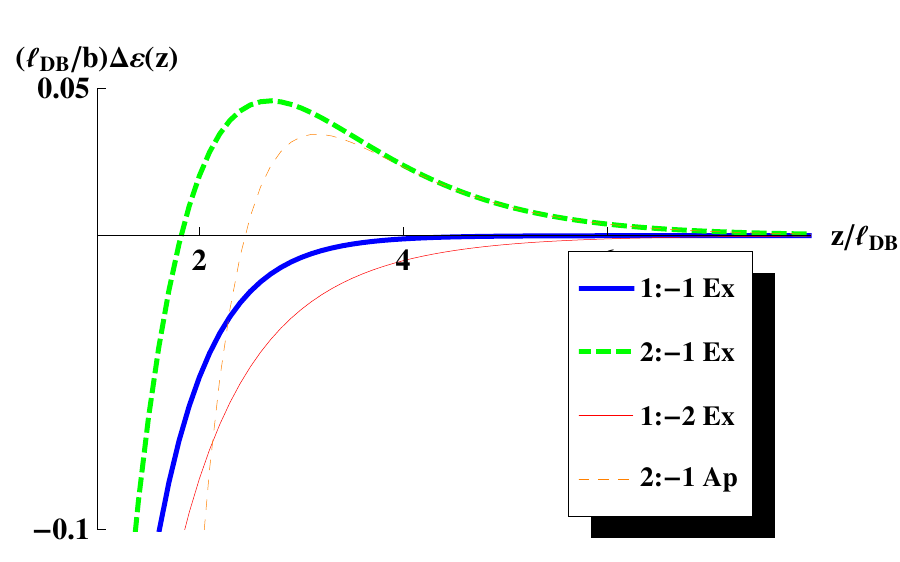}
	\caption{(Color online) {\bf Far Field}   behaviors of the correlation energy in various kinds of electrolyte, near a highly charged surface (i.e., $z_0 \ll 1$).  Blue thick solid line, $1:-1$; green thick dashed line, $2:-1$; red thin solid line, $1:-2$. These curves have been plotted using the \emph{exact} expressions Eqs.~(\ref{self-energy-1}), (\ref{Delta_E_2:1_app}) and (\ref{eq:exact_2:1_infinite_charge_form}) for an infinitely charged plate. Correction due to the finiteness of surface charge density is negligible in the far-field.  For comparison, we  also plot the leading-order far-field approximation for the $2:-1$ electrolyte (orange thin dashed line), given by Eq.~(\ref{eq:self_energy_positive_plate_bare_farfield}). This approximation becomes highly accurate for $z > 4\debye$.}
  \label{fig:farfield}
\end{figure}

\vspace{2mm}
\textbf{The far-field region}.  The far-field expansion of $\self^\infty(z)$ up to the order of $e^{- 2 z}$ is:
\ba
\label{eq:self_energy_positive_plate_bare_farfield}
 \self^\infty(z)
 &=&
\frac{\coupling}{8 \pi}
\bigg( 3 \ln (3)\,  e^{-z}
+ 6  \Big(
- \gamma  -  \ln (108 z)
\nonumber\\
&-& \frac{5}{12 z} + \frac{1}{4z^2} - \frac{1}{z^3} \Big) e^{-2 z}
+ O(e^{-3 z}) \bigg). \quad\quad
\ea
This approximation is plotted as the orange thin dashed curve in Fig.~\ref{fig:farfield}, together with the exact result Eq.~(\ref{Delta_E_2:1_app}).  One can see that they agree with each other well only for $z > 4 \ell_{\rm DB}$.  The most salient feature of this far-field expansion is that it decays as $e^{-z}$ at the leading-order, like the mean potential.  The implication of this result will be discussed  in Sec.~\ref{sec:failure-PB}.  Note also that the leading-order far-field asymptotics is positive, whereas the leading-order near-field asymptotics Eq.~(\ref{eq:bare_self_energy_positive_plate}) is negative.  Therefore the correlation energy must change sign in the intermediate region.  A plot of the full result (green thick dashed curve) in Fig.~\ref{fig:farfield} shows that the change of sign occurs at $z \approx 1.8 \, \ell_{\rm DB}$.

\subsubsection{Finite Surface Charge Density}
For finite surface charge density, the correction to the correlation potential can also be obtained in a way similar to the case of the $1:-1$ electrolyte.  The leading-order result is displayed in Appendix~\ref{app:integral:2:1_positive_plate-sub}.   The near-field asymptotics of is identical to that in the case of a $1:-1$ electrolyte, Eq.~(\ref{finite-near-field}).  Expansion in terms of $z_0$ breaks down in the extreme near-field region, where $ 0 < z-z_0 \ll z_0$.  For an asymptotic analysis valid in the extreme near-field region, see Sec.~\ref{sec:universality}.  Finally, the leading-order far-field asymptotics of Eq.~(\ref{delta_e_2:1}) is
\be
\self(z)
 =  \frac{3\theta\,\coupling z_0^3}{4\pi}
 \left( \frac{1}{z} e^{-2z} +  O(e^{-3z})
 \right),
 \label{delta_e-finite_far:2:1}
\ee
which is negligibly small compared with the zeroth-order result Eq.~(\ref{eq:self_energy_positive_plate_bare_farfield}).

\subsection{$1:-2$ electrolyte}
\label{sec:asymmetric_negative}

In $2:-1$ electrolyte ($z > z_0$), the Green's function satisfies the following equations:
\begin{subequations}
\ba
&&
- \frac{d^2}{dz^2} G(z,z';\kv)
+ \left[ k^2 + \frac{1}{3} \left( e^{2\,\Psi} + 2 \, e^{-\Psi}
\right) \right]G(z,z';\kv)
\nonumber \\
&& \quad \quad \quad \quad \quad \quad
\quad \quad \quad \quad \quad
 = \coupling
\, \delta(z-z') \quad \mbox{($z > z_0$)},
\nonumber\\
\label{Green-equation-1:2} \\
&& \left( -\frac{d^2}{dz^2} + k^2 \right) G(z, z^\prime; \kv)
= 0
\quad\quad\quad\quad \mbox{ ($z < z_0$)},
\nonumber\\
\ea
\end{subequations}
where the mean field potential $\Psi(z)$ is given by Eq.~(\ref{PBsolution-2:1-}).
Two independent homogeneous solutions $\phi_+(z)$ and $\phi_-(z)$ to Eq.~(\ref{Green-equation-1:2})  are given by:
\begin{subequations}
\label{eq:phi2:1negative}
\ba
\phi_+(z) &=&
- \frac{e^{\lambda z}}{(2 \lambda - \sqrt{3})}
\Bigg[
(\lambda - 1) (2 \lambda - 1)
\\
&-& \frac{6  e^{-z_1}
\left(
\lambda -1
 + (2 \lambda -1) e^{-z_1}
-  \lambda \, e^{-2 z_1}
 \right)}
   {(1+ e^{-z_1})(1 - 4 e^{-z_1} +  e^{-2z_1})}
\Bigg],
\nonumber\\
\phi_-(z) &=& \,\,\,
\frac{e^{- \lambda z}}{(2 \lambda + \sqrt{3})}
\,\,
\Bigg[
(\lambda + 1) (2 \lambda +1)
\\
&+& \frac{6  e^{-z_1}
\left(
\lambda + 1
 + (2 \lambda +1) e^{-z_1}
-    \lambda \,e^{-2 z_1}
 \right)}
   {(1+ e^{-z_1})(1 - 4 e^{-z_1} +  e^{-2z_1})}
\Bigg],
\nonumber
\ea
\end{subequations}
where $z_1 = z + \ln(2+ \sqrt{3})$, and $\lambda \equiv \sqrt{1+k^2}$.  These solutions exhibit the near-field asymptotics Eqs.~(\ref{asymp-phi-phi'}) and the far-field asymptotics Eqs.~(\ref{phi_pm-far}), as we demanded earlier.  The Wronskian is easily calculated using Eq.~(\ref{eq:wronskian}):
\be
\label{eq:wronskian_negative_plate}
W(\phi_+, \phi_-) = \frac{2 \lambda
\left(4 \lambda^4-5 \lambda^2+1\right)}
{(4 \lambda^2 -3)}
\ee
To obtain the F-transformed Green's function, we substitute Eqs.~(\ref{eq:phi2:1negative}), Eq.~(\ref{eq:wronskian_negative_plate}) into Eqs.~(\ref{delta-factor-def-0}), (\ref{eq:phiL_general}), and (\ref{eq:green_function_general}).

\vspace{3mm}
\subsubsection{Infinite Surface Charge Density}


The F-transformed correlation potential for the case of infinite surface charge density is:
\vspace{2mm}
\ba
\label{eq:bare_self_energy_negative_plate}
&& \delta \chi^\infty(z,z; \kv) =
G(z,z;\kv) - \lim_{z\rightarrow\infty} G(z,z;\kv)
\label{selfenergy-k-2:1-}\\
&=&
\frac {(4 \lambda^2 -3)\coupling
 \phi_-(z) \left( - \phi_+(z) +  \phi_-(z) \right)}
{2 \lambda \left(4 \lambda^4-5 \lambda^2+1\right)}
 -  \frac{g}{2 \lambda},
\nonumber
\ea
with $\phi_{\pm}(z)$ defined in Eqs.~(\ref{eq:phi2:1negative}).
To obtain the correlation potential in real space, we integrate $\chi^\infty(z,z; \kv)$ over wave vector $\kv$.  The main steps of the calculation as well as the full results (very complicated) are displayed in Appendix~\ref{app:integral:2:1_negative_plate}.  The full result is also plotted in Fig.~\ref{fig:farfield} in the far-field range.  Here we present  the near-field and far-field asymptotic behaviors.

\textbf{The near-field region}.
The near-field expansion of $\self^\infty(z)$ is given by
\ba
\label{eq:bare_self_energy_negative_plate_nearfield}
 \self^\infty(z)
&=& \frac{g}{8\pi} \bigg(
-\frac{3}{2z} + 1 - \frac{2 z}{9} + O(z^3) \bigg)
\ea
The first two terms of this series are identical to the corresponding terms of the two cases we discussed previously.

\textbf{The far-field region}.
 The far-field expansion of $\self^\infty(z)$ up to the order of $e^{- 2 z}$ is given by
\begin{widetext}
\ba
 \self^\infty(z)
&=&
\frac{\coupling}{8 \pi}
\bigg( - {3}(2-\sqrt{3})  \ln (3)\, e^{-z}
+\left(
   6 (-7 + 4\sqrt{3}) (\gamma + \ln(108z))
   +\frac{1}{2 z}( -323 + 188 \sqrt{3})
   \right)\, e^{-2z}
 + O(e^{- 3 z}) \bigg).
 \nonumber\\
\label{eq:self_energy_negative_plate_bare_farfield}
\ea
\end{widetext}
We see that the leading-order term decays as $e^{-z}$, but with a negative prefactor, c.f. Eq.~(\ref{eq:self_energy_positive_plate_bare_farfield}).

\begin{figure}
	\centering
		\includegraphics[width=8.5cm]{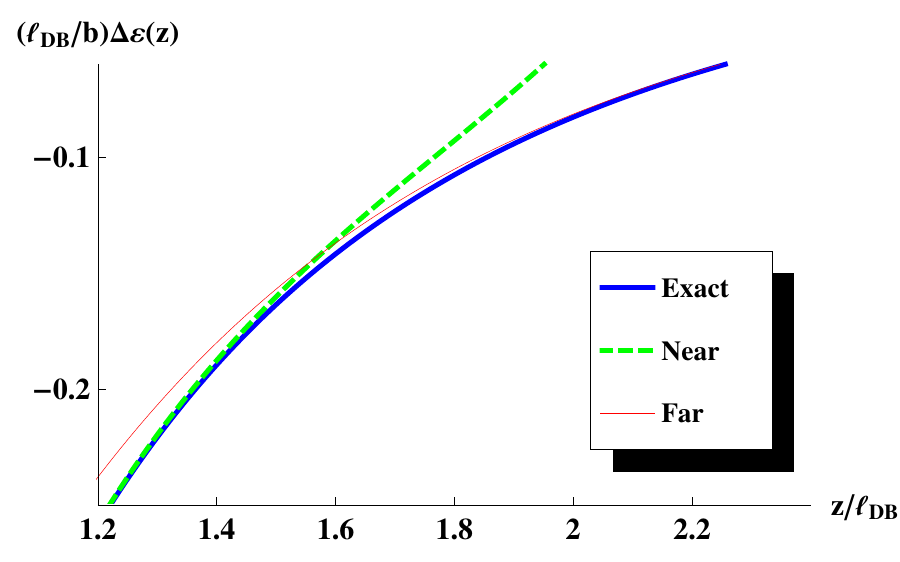}
	\caption{(Color online) {\bf 1:-2 electrolyte} \,\,
	Plots of the near-field approximation (green, thick, dashed line) up to the order of $z^8$ and far-field approximation (red, thin, solid line) up to the order of $e^{-4z}$, as well as the exact form (blue, thick, solid) [cf. Eq.~(\ref{eq:exact_2:1_infinite_charge_form})] of the correlation energy $\self(z)$ of an ion near a positively charged plate in a $1:-2$ electrolyte.  As we can see from the figure, the exact correlation energy overlaps with its near-field approximation for $z < 1.5\,\debye$, and overlaps with its far-field approximation for $z > 1.8\,\debye$.  The near  and far-field approximations at $z = 1.6\,\debye$ are accurate to within $4\%$.}
  \label{fig:matching_negative_plate}
\end{figure}

It turns out that  neither Eq.~(\ref{eq:bare_self_energy_negative_plate_nearfield})  nor Eq.~(\ref{eq:self_energy_negative_plate_bare_farfield}) is a good approximation in the intermediate region $z \sim 1$.  As shown in Fig.~\ref{fig:matching_negative_plate}, in order to achieve a moderately good matching (with error less than $4\%$), we need to go to the orders of $z^8$ in the near-field and to the order of $e^{-4 z}$ in the far-field.  These longer asymptotic expansions, together with the exact expression for $ \self^\infty(z)$, are given in Appendix~\ref{app:integral:2:1_negative_plate}.

\subsubsection{Finite Surface Charge Density}
For finite surface charge density, the correction $\delta\chi(z,z;k)$ can again be obtained using Eq.~(\ref{delta_G-finite}).  As in the previous two cases, we can expand in terms of $z_0$ to the leading order, and integrating over $\kv$, find the correction to the correlation energy.  The result is however too complicated to be exhibited here.  We will therefore only discuss its asymptotic behaviors here.
The near-field asymptotics is again identical to that in the case of the $1:-1$ electrolyte, Eq.~(\ref{finite-near-field}).  Expansion in terms of $z_0$ breaks down in the extreme near-field region, where $ 0 < z-z_0 \ll z_0$.  For an asymptotic analysis valid in the extreme near-field region, see Sec.~\ref{sec:universality}.  Finally, the leading-order far-field asymptotics of Eq.~(\ref{delta_e_2:1}) is
\be
\self(z)
= \frac{\theta\,\coupling}{\pi} (31-17\sqrt{3}) z_0^3 z^{-1} e^{-2z} + O(z_0^3\,e^{-3z}).
\ee
which is negligibly small compared with the zeroth-order results, Eq.~(\ref{eq:self_energy_negative_plate_bare_farfield}).

\section{General case of $m:-n$ electrolytes}
\label{sec:general asymmetric}
For all three cases studied above, we have shown that the near-field behaviors of the correlation energy are the same, whereas their far-field behaviors are all different.  Hence one may very well suspect that the near-field asymptotics of the correlation energy is independent of the valences of the ions.  In this section, we shall show that this is indeed the case.  Furthermore, we shall also show that inside any asymmetric electrolyte, the correlation energy decays as $e^{-z/\ell_{\rm DB}}$ in the far-field region.  The prefactor, however, depends on the valences of counterions and coions.  We further show that Poisson-Boltzmann theory breaks down in asymmetric electrolytes, regardless of the strength of the surface charge density.

\subsection{Near Field Asmptotics Independent of Valences}
\label{sec:universality}
For a strongly charged plate, the probability that a coion is near the plate is negligible.  Therefore coions should have no influence on the near-field behaviors of the mean field potential.  This, of course, has been shown explicitly for an arbitrary $m:-n$ electrolyte~\cite{xing-ming}.  As a simple illustration of the main point, we can omit the term corresponding to coions in the nonlinear PBE (assuming again a positively charged plate):
\be
-\Psi^{\prime\prime}(z) + \frac{1}{m+n}e^{n\Psi(z)} = 0.
\ee
By defining a new potential $$\widetilde{\Psi}= n\Psi + \ln \left( \frac{n}{m+n} \right),$$ the foregoing equation can be re-written in a form that does not contain any free parameters, viz.,
\be
-\widetilde{\Psi}^{\prime\prime}(z) + e^{\widetilde{\Psi}(z)}=0.
\ee
Solving this equation we find
\ba
\widetilde{\Psi}(z) &=& \ln\left( \frac{2}{z^2} \right); \\
\Psi(z) &=& \frac{1}{n}\ln\left( \frac{2(m+n)}{n\, z^2}\right).
\label{Psi-general-near}
\ea
Hence $\widetilde{\Psi}(z)$ is independent of the valences $m,n$.  The logarithmic singularity of $\Psi(z)$ and $\widetilde{\Psi}(z)$ at $z = 0$ corresponds to an infinitely charged plate at $z = 0$.  As stated in Sec.~\ref{sec:electrostatic_bc}, for finite surface charge density, we choose the position of the interface $z_0$ such that Eq.~(\ref{Psi-general-near}) remains a near-field approximation to the mean potential regardless of the value of the surface charge density $\eta$.  This requirement determines $z_0$ as a function of $\eta$ via the interface condition Eq.~(\ref{BC_dielectric-dimless}). [Note also that $\Psi(z)$ is independent of $z$ for $z< z_0$.]  For a strongly charged surface, $z_0 \ll 1$, and we can safely use Eq.~(\ref{Psi-general-near})
as a leading-order approximation of the mean potential.  This gives
\be
z_0 = \frac{2}{n \, \eta} + O(\eta^{-2}).
\ee
By the same reasoning, we also expect that coions should have no influence on the Green's function in the near-field.  Omitting the corresponding term (which is proportional to $e^{-m \Psi}$) in  Eq.~(\ref{eq:Green_function_ODE}), and plugging in the near-field asymptotic form Eq.~(\ref{Psi-general-near}) for $\Psi(z)$, the ODE for the Green's function becomes
\be
\left(-\frac{d^2}{d z^2}+k^2 + \frac{2}{z^2} \right) \, G(z, z^\prime;\kv)
= \coupling \, \delta\left( z - z^\prime \right),
\label{eq:Green_function_ODE_general}
\ee
which is indeed independent of the valences $m, n$.  Consequently, the leading-order behavior of the correlation energy in the near-field region is the same for all electrolytes whatever the values of $m$ and $n$.  Differences emerge only at sub-leading-orders in the near-field expansions of the correlation energy, as we have seen in Eqs.~(\ref{selfenergy-nearfield}), (\ref{eq:bare_self_energy_positive_plate}), and (\ref{eq:bare_self_energy_negative_plate_nearfield}).

Equation~(\ref{eq:Green_function_ODE_general}) has two linearly independent solutions $\phi_\pm(z)$:
\begin{subequations}
\label{eq:phipm_nearfield}
\ba
&&\phi_+(z) = \frac{1}{z}(1-kz)e^{kz};
\\
&&\phi_-(z) = \frac{1}{z}(1+kz)e^{-kz}.
\label{eq:phim_nearfield}
\ea
\end{subequations}
from which we deduce the Wronskian:
\be
\label{eq:wronskian_nearfield}
W=2k^3.
\ee
For $z_0 \ll 1$,  $\phi_{\pm}(z_0) $ and $\phi_{\pm}'(z_0)$ have the asymptotics that we demanded in Eqs.~(\ref{asymp-phi-phi'}):
\be
\phi_{\pm}(z_0) = \frac{1}{z_0} + O(1), \quad
\phi'_{\pm}(z_0) = - \frac{1}{z_0^2} + O(z_0^{-1}).
\ee
Using Eqs.~(\ref{delta-factor-def-0}) and (\ref{eq:phipm_nearfield}), we find the near-field approximation to the function $\delta(\kv,z_0,\epsilon_r)$:
\be
\label{eq:delta_nearfield}
\delta(\kv,z_0,\epsilon_r) =
 \frac{k(\epsilon_r-1)(z_0^{-1}-k)+z_0^{-2}}
 {k(\epsilon_r+1)(z_0^{-1}+k)+z_0^{-2}}e^{2kz_0}
 -1.
\ee
In what follows, we analyze the asymptotics of the correlation energy in two different regions: (i)~the near-field region ($z_0 \ll z - z_0 \ll 1$) and (ii)~the extreme near-field region ($0 < z - z_0 \ll z_0$). 

\vspace{2mm}
\textbf{The near-field region}. We expand the function $\delta(\kv,z_0,\epsilon_r)$ in Eq.~(\ref{eq:delta_nearfield}) in powers of the smaller parameter $z_0$.  To the leading-order term we have
\be
\delta(\kv,z_0,\epsilon_r)
\approx
\frac{4}{3}\theta k^3 z_0^3,
\ee
where $\theta=1\,(-1/2)$ for $\epsilon_r \ll 1/z_0$ ($\epsilon_r \gg 1/z_0$). Using the above result and Eqs.~(\ref{eq:green_function_general}), (\ref{eq:phiL_general}), (\ref{eq:phipm_nearfield}), and (\ref{eq:wronskian_nearfield}), we obtain the Green's function for the bulk electrolyte, viz.,
\be
G_0(z,z;\kv)=\frac{\coupling}{2k}.
\ee
which is different from the exact result Eq.~(\ref{G0-FT}).  This difference arises due to our neglect of coions, but is of no importance in the near-field.  The F-transformed correlation potential is then
\ba
\delta \chi(z,z;\kv) &=& G(z,z;\kv)-G_0(z,z;\kv)
\nonumber\\
&=&\frac{\coupling(1+kz)\left[ (1+kz)e^{-2kz}+kz-1 \right]}{2k^3z^2}
\nonumber\\
&&+\theta \frac{2\coupling z_0^3}{3z^2}(1+kz)^2 e^{-2kz}.
\ea
The first term describes the contribution for the infinitely charged plate and the second term describes the leading-order correction from the finiteness of the surface charge density. Integrating both terms over wave vectors $\bm{k}$ yields the near-field expansion of $\self^\infty(z)$ and $\self(z)$:
\begin{subequations}
\ba
\label{eq:correlation_energy_nearfield}
\self^\infty(z)
&=& \frac{\coupling}{8\pi} \cdot \left(
-\frac{3}{2z} + {1} + O(z) \right),
\label{eq:correlation_energy_nearfield-1}
\\
\self(z)
&=&\theta \cdot \frac{3\coupling}{16\pi}\cdot
\left( \frac{ z_0}{z} \right)^3 \left( \frac{1}{z}
+ O(1) \right).
\label{eq:correlation_energy_nearfield-2}
\ea
\end{subequations}
These same results have been derived for all three cases analyzed previously
[see  Eqs.~(\ref{selfenergy-nearfield}),  (\ref{finite-near-field}), Eqs.~(\ref{eq:bare_self_energy_positive_plate}), and Eqs.~(\ref{eq:bare_self_energy_negative_plate_nearfield}).]  Outside the extreme near-field region $z\gg z_0$, the correction Eq.~(\ref{eq:correlation_energy_nearfield-2}) can be neglected compared with Eq.~(\ref{eq:correlation_energy_nearfield-1}).  Hence the correlation potential is asymptotically independent of the dielectric constant of the plate.

\vspace{2mm}
\textbf{The extreme near-field region}.
In the extreme near-field region, $z - z_0 \ll z_0$, and Eq.~(\ref{eq:correlation_energy_nearfield-2}) is comparable with Eq.~(\ref{eq:correlation_energy_nearfield-1}).  Perturbation theory in $z_0$ breaks down in this region and we cannot treat $z_0$ as a small parameter.  We will therefore have to use the full expression Eq.~(\ref{eq:delta_nearfield}) for $\delta(\kv,z_0,\epsilon_r)$ to calculate the correction to the correlation energy, Eq.~(\ref{delta_e-finite}).  This is given by
\ba
\delta \varepsilon(z)
&\approx& \frac{\coupling}{4\pi}
 \int_0^\infty \frac{dk}{k^2}
{ (z^{-1} + k)^{2}
e^{-2k z}}  \times
\label{delta-e_ENF}
\\
&& \left\{
 \frac{k(\epsilon_r-1)(z_0^{-1}-k)+z_0^{-2}}
 {k(\epsilon_r+1)(z_0^{-1}+k)+z_0^{-2}}e^{2kz_0}
 -1
\right\}.\nonumber
\ea
In the extreme near-field region, the integral is dominated by the region $k \sim (z - z_0)^{-1} \gg z_0^{-1}$, and thus $\exp k z_0 \gg 1$.  Hence to obtain the leading-order result, it is legitimate to make the following approximations:
\ba
&&  z^{-1} + k \approx k, \nonumber\\
&& \frac{k(\epsilon_r-1)(z_0^{-1}-k)+z_0^{-2}}
 {k(\epsilon_r+1)(z_0^{-1}+k)+z_0^{-2}}e^{2kz_0}
 -1
\approx  \frac{1 - \epsilon_r }{1 + \epsilon_r} e^{2 k z_0}.
\nonumber
\ea
Equation~(\ref{delta-e_ENF}) then reduces to
\ba
\label{eq:correction_extreme_nearfield}
\self(z) &\approx& \frac{\coupling}{8\pi}
\cdot \frac{1 - \epsilon_r }{1 + \epsilon_r}
\int_0^{\infty} dk\, e^{-2 k (z-z_0)}
\nonumber\\
&=&  \frac{\coupling}{4\pi}
\cdot
\frac{1 - \epsilon_r }{1 + \epsilon_r}
\cdot
\frac{1}{4(z-z_0)}.
\ea
This is Eq.~(\ref{chi-extreme-near}), which describes the image charge effect arising due to the discontinuity in the dielectric constant.  In the extreme near-field, $z - z_0 \ll z_0$, Eq.~(\ref{eq:correction_extreme_nearfield}) dominates  Eq.~(\ref{eq:correlation_energy_nearfield}), and hence the correlation energy is dominated by the dielectric discontinuity.

\subsection{Far-field asymptotics depends on valences}
\label{sec:farfield}
In this section, we show that for arbitrary {\em asymmetric} $m:-n$ electrolytes ($m \neq n$), the correlation energy decays as $e^{-z}$ in the far-field region, with a prefactor that depends on the valences of counterions and coions.

It is sufficient to prove this result for the case of a plate with infinite surface charge density ($z_0 = 0$) located at the origin.  As was demonstrated in Ref.~\cite{xing-ming}, the mean-field potential can be expanded into the following far-field asymptotic series:
\be
\Psi(z) = \sum_{\ell=1}^{\infty} c_\ell \, e^{-\ell z}.
\ee
By substituting this back into the PBE~(\ref{NLPB-dimless}), and comparing coefficients order by order, all higher-order coefficients $c_k$ for $k \geq 2$ can be determined as functions of $c_1$.  For the three cases studied above, $c_1$ is exactly known:
\ba
\begin{array}{ll}
c_1 = 4 & 1:-1,\\
c_1 = 6 & 2:-1,\\
c_1 = 6 (2 - \sqrt{3}) \quad\quad& 1:-2.
\end{array}
\label{c_1-values}
\ea
For other types of electrolyte, $c_1$ can be approximately calculated.  Detailed discussions can be found in Ref.~\cite{xing-ming}.

Using the far-field expansion for $\Psi(z)$, the equation for the Green's function, Eq.~(\ref{eq:green_function_z}), can be similarly expanded:
\begin{widetext}
\be
\label{eq:series_ODE}
\left( -\frac{d^2}{dz^2} + \lambda^2
+ (n-m)\sum_{\ell=1}^\infty c_\ell e^{-\ell z}
+ \frac{1}{2}(m^2 + n^2 - m n)
\sum_{\ell, \ell^\prime = 1}^\infty c_\ell\,
 c_{\ell^\prime} e^{-\left( \ell + \ell^\prime \right) z}
+ \cdots
 \right)
G(z,z^\prime; \kv)
= \coupling \, \delta(z-z^\prime).
\ee
\end{widetext}
Terms that are ignored are at least of the order of $e^{-3z}$ in the far-field, and therefore can be ignored for our purpose.  The two homogeneous solutions $\phi_+(z)$ and $\phi_-(z)$  can also be expanded into the following Frobenius series:
\begin{subequations}
\label{far-expansions-phi_pm}
\ba
\phi_-(z)&=&e^{-\lambda z}
\bigg( 1 + \sum_{j=1}^\infty a_j\, e^{-j z} \bigg);
\\
\phi_+(z)&=&e^{\lambda z} \,\,
\bigg( 1 + \sum_{j=1}^\infty b_j\, e^{-j z} \bigg).
\ea
\end{subequations}
Since these series are applicable only in the far-field, we have no knowledge about the near-field behaviors of $\phi_{\pm}(z)$ at all.    By substituting these series into the homogeneous version of the ODE Eq.~(\ref{eq:series_ODE}), and equating terms order by order in powers of $e^{ - z }$, we can obtain values of the coefficients $a_j$ and $b_j$. For our purpose, it suffices to determine the first coefficient $a_1$ and $b_1$ for each function:
\begin{subequations}
\label{eqn-a-b}
\ba
a_1 = \frac{n-m}{1+2\lambda} \, c_1;
\\
b_1 = \frac{n-m}{1-2\lambda} \, c_1.
\ea
\end{subequations}

Equation~(\ref{eq:series_ODE}) has a Sturm-Liouville form, and therefore its Green's function can be written as the following standard form:
\be
\label{eq:G}
G(z,z^\prime; \kv) = -\frac{\coupling\,\left(\phi_+(z)
+\leftcoeff\, \phi_-(z)\right)\phi_-(z^\prime)}{W},
\ee
$W$ is the Wronskian formed by $\phi_{\pm}(z)$:
\be
W = \phi_+ \phi_-^\prime - \phi_+^\prime \phi_-.
\ee
For a Sturm-Liouville system in the form of Eq.~(\ref{eq:series_ODE}), the Wronskian is known to be independent of $z$. Therefore we need to calculate it only in the limit $z \rightarrow \infty$, and $\phi_{\pm} \sim e^{\pm \lambda z}$. This gives us  $W=-2\lambda$.

The coefficient $\leftcoeff$ is to be determined by fixing the boundary condition on the plate.  This can not be done, because our far-field expansions Eqs.~(\ref{far-expansions-phi_pm}) are not valid in the near-field region.  Fortunately enough, we are interested in only the leading-order far-field behaviors of the Green's function, and that turns out to be independent of the coefficient $\leftcoeff$.  Substituting Eqs.~(\ref{far-expansions-phi_pm})  into Eq.~(\ref{eq:G}), and setting $z=z^\prime$, we find that the leading-order approximation of the Green's function is given by
\be
G(z,z; \kv) \approx \frac{\coupling}{2 \lambda}
\left( 1+(a_1+b_1)e^{ - z} \right)
+ O(e^{-2 z}, e^{-2 \lambda z}).
\ee
Note that we have neglected a contribution proportional to $e^{-2\lambda z}$. As $\lambda = \sqrt{1+k^2} \geq 1$, the latter is indeed subdominant in the far-field region.  The $k$-dependent correlation potential is now given by
\be
\delta \chi(z,z; \kv) = \frac{\coupling(a_1 + b_1) \, e^{- z }}{2 \lambda} + O(e^{- 2 z}),
\ee
from which we deduce the correlation energy:
\ba
\label{eq:asymmetric}
\self(z) &=&
\frac{\coupling}{2}
\int_0^{\infty} \frac{d^2k}{(2 \pi)^2}\chi(z,z; \kv)
\nonumber\\
&=& \frac{\coupling}{2} \cdot \frac{1}{2\pi}
\int_{1}^{\infty}d\lambda \,  \frac{1}{2}  (a_1 + b_1) \, e^{- z }
\nonumber\\
&=& \frac{\coupling}{2} \cdot \frac{1}{2\pi}
\int_{1}^{\infty}d\lambda \, \frac{(n-m)
 \, c_1}{1-4\lambda^2}e^{- z }
\nonumber\\
&=& \frac{\coupling}{16\pi} \ln (3) \,(m - n) \, c_1e^{- z }.
\ea
Combined with Eq.~(\ref{c_1-values}), we see that Eq.~(\ref{eq:asymmetric}) gives the same leading-order far-field asymptotics for the correlation energies in a $2:-1$ electrolyte [cf. Eq.~(\ref{eq:self_energy_positive_plate_bare_farfield})] and  an ion in a $1:-2$ electrolyte [cf. Eq.~(\ref{eq:self_energy_negative_plate_bare_farfield})].
Therefore, we conclude that inside any asymmetric electrolyte, the correlation energy in the far-field indeed decays as $e^{-z}$.

\subsection{Breakdown of Perturbation Theory in Asymmetric Electrolytes}
\label{sec:failure-PB}

To see whether fluctuation correlation effects are important in the far-field region, let us substitute the value of the correlation energy calculated in Eq.~(\ref{eq:asymmetric}) into the FCPBE~(\ref{eq:mpbe}), and try to solve perturbatively for the average local potential.  The dimensionless version of the FCPBE is given by
\be
\label{eq:modified_PBE}
- \Delta \Psi + \frac{1}{m+n} \left(
e^{n \Psi - n^2 \self}
- e^{-m \Psi - m^2 \self}
\right) = 0.
\ee
Perturbation theory can be performed by treating $\coupling$ as a small parameter.  Furthermore, since we have calculated the correlation energy only up to first order in $\coupling$, perturbation theory for Eq.~(\ref{eq:modified_PBE}) is reliable only up to the same order.  Hence let us write  $\Psi \approx \Psi_0 + \Psi_1$, where $\Psi_0$ is the mean field solution and $\Psi_1$ is of the order of $\coupling$. \footnote{It can be shown that this first order perturbation analysis in $\coupling$ is equivalent to the one-loop approximation in the Sine-Gordon field theory representation of the original Coulomb many body problem. } Equating terms of the same order in $\coupling$ shows that $\Psi_1$ satisfies the following linear inhomogeneous equation:
\be
-\Delta \Psi_1 + \frac{1}{m+n}
\left( n \, e^{n \Psi_0} + m \, e^{-m \Psi_0}
\right) \Psi_1
= -(m-n) \self.
\ee
Again, we consider plate geometry and employ the far-field asymptotic form of $\self$ from Eq.~(\ref{eq:asymmetric}):
\be
- \Psi_1''(z) + \Psi_1(z)
= -\frac{\coupling}{16\pi} \ln (3) \,(m - n)^2 \, c_1e^{- z }.
\ee
This yields the following asymptotic form of $\Psi_1(z)$ for the far-field region:
\be
\Psi_1(z) = -\frac{\coupling}{32\pi} \ln (3) \,(m - n)^2 \, c_1\, z \,e^{-z}
+ b \, e^{-z},
\ee
where $b$ is an integration constant to be determined by boundary conditions.  The first term is {\em secular}, and becomes larger than the zeroth-order approximation $\Psi_0(z) \sim c_1 \, e^{- z}$ for sufficiently large $z$, implying the breakdown of perturbation theory.  It is thus inconsistent to treat the correlation energy as a perturbation, even in the far-field region: both correlation energy and mean field energy must be treated on equal footing, e.g., within the self-consistent Gaussian approximation.   We shall study this theory in a future presentation~\cite{ding}.

\vspace{1mm}
\section{Conclusion}
\label{sec:conclusion}
One of the most salient features of the nonlinear Poisson-Boltzmann theory is that the electrostatic potential $\Phi$ (at nonzero distance from the charged surface) remains finite even if the surface charge density $\sigma$ becomes infinitely large, as has been shown in previous works (see, e.g., \cite{alexander}, \cite{Tellez:2011fk}, \cite{xing-ming}, and \cite{xing-two-plate}).  In Ref.~\cite{xing-ming}, the renormalized charge density $\sigma_R$ of a charged plate was obtained as an asymptotic series of $\sigma$ for various cases of the $m:-n$ asymmetric electrolyte.

In this work, we have proceeded one step further by studying the correlation potential of a test ion near a strongly charged plate inside an $m:-n$ electrolyte, and have obtained the following general results:

(1) For an infinitely charged plate, the correlation potential is independent of the dielectric constant $\epsilon_1$ of the plate.
(2) For a strongly (but finitely) charged plate, the correlation potential depends on $\epsilon_1$, but this dependence becomes negligible when the distance $\Delta z$ between the test ion and the plate is much larger than $\mu$.
(3) If the distance to the plate is much smaller than $\mu$, the correlation potential is dominated by the image charge effect arising from the discontinuity of permittivity across the interface, but is independent of the type of electrolyte.
(4) In the region $\mu \ll \Delta z \ll \ell_{\rm DB}$, where $\ell_{\rm DB}$ is the Debye length, the correlation potential can be described by a point-like image charge with strength $q_{\rm im} = - 3 q$ at the mirror point.  This result depends neither on the permittivity of the plate nor on the type of electrolyte.
(5) The far-field ($\Delta z \gg \ell_{\rm DB}$) asymptotics of the correlation potential explicitly depends on the valences of ions, but is independent of the permittivity of the plate.
(6) More importantly, for any {\em asymmetric} electrolyte ($m \neq n$), the correlation potential decays as $\exp (- z/\debye)$ in the far-field region, i.e., with the same decay width as the mean field potential energy.  This implies the breakdown of perturbative calculations of the average potential, even for small (but non-zero) values of the coupling parameter $\coupling$.

We shall explore the consequences of these results further in future presentations.

The authors thank NSFC (Grants No. 11174196 and No. 91130012) for financial support.


\begin{thebibliography}{99}
\bibitem{footnote1}
Some of these charges may be fixed, so that the average potential $\Phi(\rv)$ is generically nonzero.  Eqs.~(\ref{H_N-def}) and (\ref{varphi-def}) apply to the case where there is no dielectric discontinuity in the medium, but it is straightforward to include effects from image charges if a region of different dielectric permittivity is present. For example, $v(\xv, \yv)$ would need to be modified to include both the image charge potential as well as the Coulomb potential.
\bibitem{footnote3}
In reality, as long as the plate thickness is much larger than the Debye length, it can be effectively approximated by infinity.
\bibitem{footnote4}
In reality, the plate must have finite thickness.  The mean field potential is constant if the two sides of the plate carry identical surface charge densities.
\bibitem{footnote5}
Detailed calculation shows that $\delta(\kv,z_0,\epsilon_r)$ is of the order of $z_0^3$.  We do not need this refined result here.
\end{thebibliography}

\appendix

\section{Green's Function is Independent of Choice of $\phi_+(z)$}
\label{app:gauge-transform}
In Sec.~\ref{sec:green_function}, we have defined a homogeneous solution $\phi_+(z)$ to Eq.~(\ref{eq:green_function_z})  that is exponentially increasing as $e^{\lambda z}$ for large $z$.  This requirement however determines $\phi_+(z)$ only up to a linear superposition of $\phi_-(z)$.   The Green's function $G(z,z';\kv)$, on the other hand, must be independent of the choice of $\phi_+(z)$.  Here we show this independence.  Let us make the following ``gauge transformation'':
\be
\phi_+(z) \rightarrow \phi_+(z) + a\, \phi_-(z).
\label{Gauge-transform}
\ee
We need to prove only that the Green's function remains invariant under this transformation.

For $z<z_0$, $G(z,z';\kv)$ is given by the first line of Eq.~(\ref{eq:green_function_general}), and does not depend on $\phi_+(z)$.
It is therefore manifestly invariant under the transformation Eq.~(\ref{Gauge-transform}).  For $z > z_0$, $G(z,z';\kv)$ is given by the second line of Eq.~(\ref{eq:green_function_general}), which depends on $\phi_+(z)$ through the Wronskian and through $\phi_L(z)$.  The Wronskian Eq.~(\ref{eq:wronskian}) is clearly invariant under the transformation Eq.~(\ref{Gauge-transform}).  The function $\phi_L(z)$ is defined by Eq.~(\ref{eq:phiL_general}).  Using Eq.~(\ref{delta-factor-def-0}), it can be rewritten as
\be
\phi_L(z) = - \phi_+(z) +  \frac{k \epsilon_r \phi_+(z_0)
- \phi_+^\prime(z_0)}{k \epsilon_r \phi_-(z_0)
 - \phi_-^\prime(z_0)} \, \phi_-(z),
\ee
which is also invariant under the transformation Eq.~(\ref{Gauge-transform}).  Hence the Green's function Eq.~(\ref{eq:green_function_general}) is independent of the choice of $\phi_+(z)$.

\section{Calculation of correlation energy for $1:-1$ electrolyte}
\label{app:integral:1:-1}
In this appendix we calculate the integral Eq.~(\ref{chi-integral}), with $\chi^\infty(z,z;\kv)$ given by Eq.~(\ref{self-energy-01}).
This integral is complicated by the fact that both the denominator and the numerator
of  Eq.~(\ref{self-energy-01}) vanish at $\lambda = 1$.  To resolve this issue, we make a variable transformation as follows:
\ba
u = \lambda - 1 = \sqrt{k^2 + 1} -1.
\ea
Equation~(\ref{chi-integral}) can then be rewritten into the following form:
\ba
\frac{8 \pi}{\coupling}  \self^\infty(z) &=&
 {\rm csch}^2 (z) \cdot
 \int_0^{\infty} \frac{ (e^{- 2 u z} - 1)du}  { u(u+2)}
\\
 &+&
  e^{-2 z} \int_0^{\infty}
  \left(
  1 + \frac{2}{2+u} \coth (z)
  \right)e^{- 2 u z}  du. \nonumber
\ea
Each integral in the right hand side converges separately.  The final result is displayed in Eq.~(\ref{self-energy-1}).

\section{Calculation of correlation energy for $2:-1$ electrolyte}
\label{app:integral:2:1_positive_plate}
To obtain the correlation energy for the case $z_0 = 0$ [which we denote by the symbol $\self^\infty(z)$], we need to integrate $\chi^\infty(z,z; \kv)$ in Eq.~(\ref{selfenergy-k-2:1}) over wave vectors $k$.  Note however that this integration is complicated by the vanishing of the denominator as $\lambda - 1$ when $\lambda \rightarrow 1$ (which corresponds to the limit $k \rightarrow 0$, as $\lambda = \sqrt{1+ k^2}$). On the other hand, we know that the integral is convergent [as we have already subtracted the truly divergent part $G_0(z, z; \kv)$]. Thus the pole at $\lambda = 1$ in the denominator must be canceled by a corresponding pole in the numerator.  To ensure that our integration is convergent, we should explicitly isolate the pole in the numerator.  We therefore adopt the following procedure.  We first define the following functions:
\ba
\label{eq:alpha_beta_functions}
\alpha(z,\lambda) &\equiv& - \phi_+(z) \phi_-(z)
-  (4 \lambda^4 - 5 \lambda^2 +1),
\nonumber\\
\beta(z,\lambda) &\equiv& \phi_-(z)^2 e^{2(\lambda -1 )z},
\nonumber\\
\delta \alpha(z,\lambda) &\equiv& \alpha(z,\lambda) - \alpha(z,1),
\nonumber\\
\delta \beta(z,\lambda) &\equiv& \beta(z,\lambda) - \beta(z,1).
\ea
Here, the functions $\phi_{\pm}(z)$ are defined as in Eqs.~(\ref{eq:phi2:1positive}), and the corresponding Wronskian is given by
\be
W = \frac{1}{2\lambda}
(4\lambda^4-5\lambda^2+1).
\ee
Using Eqs.~(\ref{G_G0_relation}), (\ref{eq:green_function_general}) and (\ref{eq:phiL_general}), we can write the wave vector dependent correlation potential $\chi^\infty(z,z;k)$ as
\be
\delta \chi^\infty(z,z;\kv) =
\coupling \frac{- \phi_+(z) \phi_-(z)+\phi_-(z)^2}
{(4\lambda^4-5\lambda^2+1)/2\lambda} - \frac{\coupling}{2 \lambda}
\ee
This gives Eq.~(\ref{selfenergy-k-2:1}).
The superscript $\infty$ indicates that we are considering the case of an infinite surface charge density, i.e., $z_0=0$, which means that $\delta(\kv,z_0,\epsilon_r)=0$ [cf. Eq.~(\ref{delta-factor-def})].
Now we apply Eqs.~(\ref{eq:alpha_beta_functions}) to rewrite the correlation potential $\chi^\infty(z,z;k)$ as follows:
\ba
\delta \chi^\infty(z,z;\kv)
&=&
\coupling \frac{\alpha(z,\lambda) + \beta(z,\lambda) \,e^{-2 (\lambda -1) z}}{2\lambda(4\lambda^4-5\lambda^2+1)}
 \label{eq:chi_infty_reexpressed} \\
&=&
\frac{\coupling}{2\lambda(4\lambda^4-5\lambda^2+1)}
\Big(  \alpha(z,1) + \delta\alpha(z,\lambda)
\nonumber\\
 &+&
 \left. \beta(z,1) e^{-2 (\lambda -1) z}
 + \delta \beta(z,\lambda) e^{- 2 (\lambda -1) z} \right)
\nonumber
\ea
It is straightforward to compute the following quantities:
\begin{subequations}
\label{eq:C3}
\ba
\alpha(z,1) &= & - \beta(z,1) =
-\frac{36 e^{2 z} \left(e^z+1\right)^2}{\left(-3
   e^z+3 e^{2 z}+e^{3 z}-1\right)^2},
  \\
\delta \alpha(z,\lambda) &=&
\frac{12 e^z \left(-4 e^z-6 e^{2 z}-4 e^{3
   z}+e^{4 z}+1\right) \left(\lambda
   ^2-1\right)}{\left(-3 e^z+3 e^{2 z}+e^{3
   z}-1\right)^2},
   \nonumber\\
  \\
\delta \beta(z,\lambda) &=&
   \frac{e^{-2 z} (\lambda
   -1)}{\left(e^z-1\right)^2 \left(4 e^z+e^{2
   z}+1\right)^2}
   \sum_{m = 0}^6 p_m(\lambda) \, e^{m z},
   \nonumber\\
   \label{eq:numerator_symmetric_electrolyte}
\ea
\end{subequations}
where $p_m(\lambda)$ are all polynomials of $\lambda$ of degree 3, defined as
\ba
p_0(\lambda) &\equiv&
 4 \lambda ^3-8 \lambda ^2+5 \lambda -1,
  \nonumber\\
p_1 (\lambda) &\equiv&
6 \left(4 \lambda ^3-4 \lambda ^2-\lambda + 1\right),
 \nonumber\\
p_2(\lambda ) &\equiv&
3 \left(4 \lambda ^3-3 \lambda -5\right),
 \nonumber\\
p_3(\lambda) &\equiv&
20 \left(-4 \lambda ^3-4 \lambda ^2+\lambda +1\right),
 \nonumber\\
p_4(\lambda) &\equiv&
3 \left(4 \lambda ^3+8 \lambda ^2+5 \lambda + 7\right),
 \nonumber\\
p_5(\lambda) &\equiv&
24 \lambda ^3+72 \lambda ^2+90 \lambda +78,
 \nonumber\\
p_6(\lambda) &\equiv&
4 \lambda ^3+16 \lambda ^2+29 \lambda +35.
\ea
These quantities enable us to write Eq.~(\ref{eq:chi_infty_reexpressed}) as follows:
\ba
\label{eq:C5}
\delta  \chi^\infty(z,z;\kv)
&=& \frac{\coupling} {2\lambda(4\lambda^4-5\lambda^2+1)}
\\
&& \Big( \delta \alpha(z,\lambda) +
\alpha(z,1) \left( 1 - e^{-2 (\lambda -1) z} \right)
\nonumber\\
&+& \delta \beta(z,\lambda)
 e^{- 2 (\lambda -1) z}
\Big).
\nonumber
\ea
In this form, we easily see that each of the terms in the numerator vanishes as $\lambda \rightarrow 1$, thus exactly canceling the pole $\lambda - 1$ in the denominator.

The correlation energy $\self^\infty(z)$ is given by the following, viz.,
\ba
\self^\infty(z) &=& \frac{1}{2}\int\frac{d^2k}{(2\pi)^2}\chi^\infty(z,z;\kv)
\nonumber\\
&=& \frac{1}{4\pi}\int_1^{\infty} d \lambda\, \lambda\, \chi^\infty(z,z;\kv)
\nonumber\\
&=&
\frac{\coupling}{4\pi}\int_1^{\infty} d\lambda
\frac{1}{2(4\lambda^4-5\lambda^2+1)} \times
\nonumber\\
&&\Bigg[
\alpha(z,1) \left( 1 - e^{-2 (\lambda -1) z} \right)
\nonumber\\
&+& \delta \alpha(z,\lambda)
+ \delta \beta(z,\lambda)
 e^{- 2 (\lambda -1) z}
 \Bigg]
\label{eq:chi_infty}
\ea
In order to evaluate the integral, we make use of the following results:
\ba
&&\int_1^\infty d\lambda \frac{\lambda^2-1}{2(4\lambda^4-5\lambda^2+1)} = \frac{\ln(3)}{8};
\\
&&\int_1^\infty d\lambda \frac{1-e^{-2(\lambda-1)z}}
{2(4\lambda^4-5\lambda^2+1)} =
 \frac{1}{24}\Bigg[ 2\gamma + 2e^{4z} E_1(4z)
 \nonumber\\
&&- 4e^{3z} E_1(3z) + 4e^{z} E_1(z) + \ln\left( \frac{16z^2}{81} \right) \Bigg].
\ea
Applying these results and Eqs.~(\ref{eq:C3}), and performing the integrals over $\lambda$ from $1$ to $\infty$ in Eq.~(\ref{eq:chi_infty}), we obtain the following result for the correlation energy of a test ion in front of a plate with infinite surface charge density (the plate being positioned at $z_0=0$):
\ba
&&(4\pi/\coupling)\self^\infty(z)
=
\frac{36e^{2z} (e^z+1)^2 w(z)}{(1+3e^z-3e^{2z}-e^{3z})^2}
\nonumber\\
&+& \frac{3\ln(3) \, e^z (1-4e^z-6e^{2z}
-4e^{3z}+e^{4z})}{2(1+3e^z-3e^{2z}-e^{3z})^2}
\nonumber\\
&+&
\frac{e^{-2z}}{(e^z-1)^2 (1+4e^z+e^{2z})^2}
\sum_{m=0}^{6}q_m(z) \, e^{mz},
\nonumber\\
\label{Delta_E_2:1_app}
\ea
where the functions $w(z)$ and $q_m(z)$ ($m=0,\ldots,6$) are defined by
\ba
w(z) &\equiv& -\frac{1}{24} \Big(
- 2 \ln \left(\frac{4 z}{9}\right)
-4 e^z E_1(z)
  \nonumber\\
  &+& 4 e^{3 z} E_1(3z)-2 e^{4 z}
   E_1(4 z)-2 \gamma \Big);
   \nonumber
\ea
\ba
q_0(z) &\equiv& \frac{1}{z}+6 e^{3 z} \left(E_1(3 z)-2 e^z E_1(4 z)\right);
   \nonumber\\
   q_1(z) &\equiv& 3 \left(\frac{1}{2 z}-2 e^{4 z} E_1 (4 z)\right);
\nonumber\\
q_2(z) &\equiv& \frac{3}{4} \left(\frac{1}{z}-2 e^z E_1(z)+4 e^{3 z}
   E_1(3 z)-4 e^{4 z} E_1(4 z)\right);
  \nonumber\\
  q_3(z) &\equiv& -\frac{5}{z};
\nonumber\\
q_4(z) &\equiv& \frac{3}{4} \left(\frac{1}{z}+4 e^z E_1(z)-6 e^{3 z}
   E_1(3 z)+4 e^{4 z} E_1(4 z)\right);
   \nonumber\\
q_5(z) &\equiv& 3 \left(\frac{1}{2 z}+4 e^z E_1(z)-4 e^{3 z} E_1(3 z)+2 e^{4 z} E_1(4 z)\right);
   \nonumber\\
q_6(z) &\equiv& \frac{1}{4 z}+\frac{9 e^z E_1(z)}{2}-6 e^{3 z} E_1(3 z)+3 e^{4 z} E_1(4 z).
\nonumber
\ea
The expression for the correlation energy simplifies mathematically and becomes physically transparent in the near and far-field asymptotic limits.  The asymptotic forms are presented in Eqs.~(\ref{eq:bare_self_energy_positive_plate}) and (\ref{eq:self_energy_negative_plate_bare_farfield}).

\vspace{2mm}
\subsection{Correction due to Finiteness of Surface Charge Density}
\label{app:integral:2:1_positive_plate-sub}
For finite surface charge density, the correction $\delta\chi(z,z;k)$ can again be obtained using Eq.~(\ref{delta_G-finite}).  Expanding in terms of $z_0$ to the leading-order, and integrating over $\kv$, we find that the correlation to the correlation energy is
\be
\self(z) \approx \frac{\theta\,\coupling  z_0^3}
{192\pi} \cdot \frac{128 \,
 e^z E_1(2z)\sinh^6(z/2) \, z^4 + g(z) \, e^{-z}}
 {(e^z-1)^2 (\cosh(z)+2)^2 \, z^4} ,
\label{delta_e_2:1}
\ee
where, again, $\theta = 1$ $(-1/2)$ for $\epsilon_r z_0 \ll 1 $ ($ \gg 1 $), and
\ba
g(z) &\equiv& 10 \left(-6 -12 z - 7 z^2 + 2 z^3 \right)
\nonumber\\
&+& 3 \left(6 + 12 z + 9 z^2 + 2 z^3
 \right) \cosh(z)
\nonumber\\
 &+&  6 \left(
 6 + 12 z + 15 z^2 + 14 z^3
 \right) \cosh(2z)
\nonumber\\
 &+& \left(
 6 + 12 z + 25 z^2 + 34 z^3
 \right) \cosh(3z)
 \\
 &+& 24 z \left(2 + 4 z + 3 z^2 \right)\sinh(2z)
 \nonumber\\
 &+& 24 z \big( 1 +2  z + 6 z^2\big) \cosh(z) \sinh(2z).
\nonumber
\ea

\section{Calculation of correlation energy for $1:-2$ electrolyte}
\label{app:integral:2:1_negative_plate}
To obtain the correlation energy, we integrate $\chi^\infty(z,z; \kv)$ over all wave vectors $\bm{k}$.  As for the $2:1$ electrolyte system with a positively charged plate, the integration is complicated by the fact that the denominator in the expression above vanishes when $\lambda = 1$.  Thus we shall also perform a procedure similar to that in the system with one positively-charged plate to isolate the pole at $\lambda = 1$ in the numerator.  We first define the following useful quantities:
\ba
\label{eq:functions}
\nu(z,\lambda) &\equiv& -\phi_+(z) \phi_-(z)
-  \frac{4(4 \lambda^4 - 5 \lambda^2 +1)}{4\lambda^2-3},
\nonumber\\
\tau(z,\lambda) &\equiv& \gamma(\lambda) \phi_-(z)^2 e^{2(\lambda -1 )z},
\nonumber\\
\delta \nu(z,\lambda) &\equiv& \nu(z,\lambda) - \nu(z,1),
\nonumber\\
\delta \tau(z,\lambda) &\equiv& \tau(z,\lambda) - \tau(z,1).
\ea
Here the functions $\phi_{\pm}(z)$ are defined as in Eqs.~(\ref{eq:phi2:1negative}), and the corresponding Wronskian has been given in Eq.~(\ref{eq:wronskian_negative_plate}).

Using Eqs.~(\ref{G_G0_relation}), (\ref{eq:green_function_general}), and (\ref{eq:phiL_general}), we can write the wave vector dependent correlation potential $\chi^\infty(z,z;\kv)$ as
\begin{widetext}
\be
\delta  \chi^\infty(z,z;\kv) =
\frac{(4\lambda^2-3)\coupling}{2\lambda(4\lambda^4-5\lambda^2+1)}
\left\{ \big(- \phi_+(z) + \phi_-(z) \big)\phi_-(z) - \frac{(4\lambda^4-5\lambda^2+1)}{4\lambda^2-3} \right\}.
\ee
We apply Eq.~(\ref{eq:functions}) to re-express the correlation potential as
\be
\label{eq:chi_infty_re}
\delta \chi^\infty(z,z;\kv) = \frac{(4\lambda^2-3)\coupling}{8\lambda(4\lambda^4-5\lambda^2+1)}
\left\{ \nu(z,1) +\delta\nu(z,\lambda) + \left( \tau(z,1) + \delta\tau(z,\lambda) \right)e^{-2(\lambda-1)z} \right\}.
\ee
It is straightforward to compute the following quantities:
\begin{subequations}
\ba
\nu(z,1) &= & - \tau(z,1) =
\frac{144 e^{2 z} \left(\left(2+\sqrt{3}\right) e^z-1\right)
   \left(\left(362+209 \sqrt{3}\right) e^z-97-56
   \sqrt{3}\right)}{\left(\sqrt{3}-2\right) \left(2+\sqrt{3}\right)^3
   \left(e^z-1\right)^2 \left(\left(5+3 \sqrt{3}\right) e^z+\left(26+15
   \sqrt{3}\right) e^{2 z}-1\right)^2},
   \\
\delta \nu(z,\lambda) &=&
-\frac{96 \left(\lambda^2-1\right) e^{2 z} \left(\left(9360+5404
   \sqrt{3}\right) \sinh (z)+7 \left(1351+780 \sqrt{3}\right) \cosh
   (z)-780 \sqrt{3}-1351\right)}{\left(2+\sqrt{3}\right)^3 \left(4
   \lambda^2-3\right) \left(e^z \left(\left(26+15 \sqrt{3}\right) e^z+5+3
   \sqrt{3}\right)-1\right)^2},
   \\
\delta \tau(z,\lambda) &=&
   \frac{e^{-2 z}}{(\lambda+\sqrt{3}/2)^2\left(e^{z_1}+1\right)^2\left(e^{2z_1}-4e^{z_1}+1\right)^2}
   \sum_{m = 0}^6 r_m(\lambda) \, e^{m z_1},
\ea
\end{subequations}
where $r_m(\lambda)$ are all polynomials of $\lambda$ of degree three, defined as
\ba
r_0(\lambda) &\equiv&
 \left(2 \lambda^2-3 \lambda+1\right)^2,
  \nonumber\\
r_1 (\lambda) &\equiv&
-6 (\lambda-1)^2 \left(4 \lambda^2-1\right),
 \nonumber\\
r_2(\lambda ) &\equiv&
3 (\lambda-1) \left(4 \lambda^3-3 \lambda-5\right),
 \nonumber\\
r_3(\lambda) &\equiv&
20 \left(4 \lambda^4-5 \lambda^2+1\right),
 \nonumber\\
r_4(\lambda) &\equiv&
3 (\lambda-1) \left(\lambda \left(4 \lambda (\lambda+2)+192 \sqrt{3}-331\right)-144
   \sqrt{3}+247\right),
 \nonumber\\
r_5(\lambda) &\equiv&
-6 (\lambda-1) \left(\lambda \left(4 \lambda (\lambda+3)+192 \sqrt{3}-321\right)-144
   \sqrt{3}+253\right),
 \nonumber\\
r_6(\lambda) &\equiv&
(\lambda-1) \left(\lambda \left(4 \lambda (\lambda+4)+576 \sqrt{3}-979\right)-432
   \sqrt{3}+755\right).
\ea
These quantities enable us to write Eq.~(\ref{eq:chi_infty_re}) as follows:
\be
\label{eq:chi_negative_plate}
\delta \chi^\infty(z,z;\kv)
=
\frac{(4\lambda^2-3)\coupling}{8\lambda(4\lambda^4-5\lambda^2+1)}
\left\{\nu(z,1) \left( 1 - e^{-2 (\lambda -1) z} \right)
+ \delta \nu(z,\lambda)
+ \delta \tau(z,\lambda)
 e^{- 2 (\lambda -1) z}
\right\}.
\ee
In this form, we easily see that each of the terms in the numerator vanishes as $\lambda \rightarrow 1$, thus exactly canceling the pole $\lambda - 1$ in the denominator.

Equation~(\ref{eq:chi_negative_plate}) leads to the following form for the correlation energy:
\be
\self^\infty(z)
=
\frac{\coupling}{16\pi}\int d\lambda
\frac{4\lambda^2-3}{2(4\lambda^4-5\lambda^2+1)}
\left(\nu(z,1) \left( 1 - e^{-2 (\lambda -1) z} \right)
+ \delta \nu(z,\lambda)
+ \delta \tau(z,\lambda)
 e^{- 2 (\lambda -1) z}
\right)
\ee
The momentum integral can be straightforwardly evaluated, and we obtain the following result for the correlation energy:
\ba
\label{eq:exact_2:1_infinite_charge_form}
(4\pi/\coupling)\self^\infty(z)
&=&
\frac{36 e^{2 z} \left(\left(2+\sqrt{3}\right) e^z-1\right)
   \left(\left(362+209 \sqrt{3}\right) e^z-97-56
   \sqrt{3}\right)\,v(z)}{\left(\sqrt{3}-2\right) \left(2+\sqrt{3}\right)^3
   \left(e^z-1\right)^2 \left(\left(5+3 \sqrt{3}\right) e^z+\left(26+15
   \sqrt{3}\right) e^{2 z}-1\right)^2}
\nonumber\\
&&-
\frac{3\ln(3) \, e^{2z} \left( \left(9360+5404 \sqrt{3}\right) \sinh (z)+7 \left(1351+780
   \sqrt{3}\right) \cosh (z)-780 \sqrt{3}-1351\right)}{\left(2+\sqrt{3}\right)^3 \left(e^z \left(\left(26+15 \sqrt{3}\right)
   e^z+5+3 \sqrt{3}\right)-1\right)^2}
\nonumber\\
&&+
\frac{e^{-2z}}{4\left(e^{z_1}+1\right)^2\left(e^{2z_1}-4e^{z_1}+1\right)^2}\sum_{m=0}^{6}s_m(z) \, e^{mz_1},
\ea
where the functions $v(z)$ and $s_m(z)$ ($m=0,\ldots,6$) are defined by
\ba
v(z) &\equiv& \frac{1}{12} \left(\ln(324z) - 4e^z E_1(z)
         + 4 e^{3 z} E_1(3 z) + e^{4 z} E_1(4 z) + \gamma \right);
   \nonumber\\
s_0(z) &\equiv& \frac{1}{z}-6 \left(2+\sqrt{3}\right) e^{3 z} E_1(3 z)-12
   \left(7+4 \sqrt{3}\right) e^{4 z} E_1(4 z)+2 \left(45+26
   \sqrt{3}\right) e^{\left(2+\sqrt{3}\right) z} E_1\left(\left(2+\sqrt{3}\right) z\right);
   \nonumber\\
s_1(z) &\equiv& -\frac{6}{z}+24 \left(7+4 \sqrt{3}\right) e^{4 z} E_1(4 z)-12
   \left(12+7 \sqrt{3}\right) e^{\left(2+\sqrt{3}\right) z} E_1\left(\left(2+\sqrt{3}\right) z\right);
\nonumber\\
s_2(z) &\equiv& \frac{3}{z}-6 \left(\sqrt{3}-2\right) e^z E_1(z)-12
   \left(2+\sqrt{3}\right) e^{3 z} E_1(3 z)-12 \left(7+4
   \sqrt{3}\right) e^{4 z} E_1(4 z)
   \nonumber\\
   &&+30 \left(3+2 \sqrt{3}\right)
   e^{\left(2+\sqrt{3}\right) z} E_1 \left(\left(2+\sqrt{3}\right)
   z\right);
\nonumber\\
s_3(z) &\equiv& \frac{20}{z}-40 \sqrt{3} e^{\left(2+\sqrt{3}\right) z} E_1
   \left(\left(2+\sqrt{3}\right) z\right)
      \nonumber\\
s_4(z) &\equiv& \frac{3}{z}+12 \left(15 \sqrt{3}-26\right) e^z E_1
   (z)+\left(324-198 \sqrt{3}\right) e^{3 z} E_1 (3
   z)+\left(84-48 \sqrt{3}\right) e^{4 z} E_1 (4 z)
   \nonumber\\
   &&+30 \left(2
   \sqrt{3}-3\right) e^{\left(2+\sqrt{3}\right) z} E_1
   \left(\left(2+\sqrt{3}\right) z\right); \\
s_5(z) &\equiv& -\frac{6}{z}+96 \left(7-4 \sqrt{3}\right) e^z E_1 (z)+96 \left(4
   \sqrt{3}-7\right) e^{3 z} E_1 (3 z)+24 \left(4
   \sqrt{3}-7\right) e^{4 z} E_1 (4 z)
   \nonumber\\
   &&-12 \left(7
   \sqrt{3}-12\right) e^{\left(2+\sqrt{3}\right) z} E_1
   \left(\left(2+\sqrt{3}\right) z\right);
   \nonumber\\
s_6(z) &\equiv& \frac{1}{z}+6 \left(31 \sqrt{3}-54\right) e^z E_1 (z)+48 \left(7-4
   \sqrt{3}\right) e^{3 z} E_1 (3 z)+\left(84-48 \sqrt{3}\right)
   e^{4 z} E_1 (4 z)
   \nonumber\\
   &&+2 \left(26 \sqrt{3}-45\right)
   e^{\left(2+\sqrt{3}\right) z} E_1 \left(\left(2+\sqrt{3}\right)
   z\right). \nonumber
\ea
The results for the asymptotic behavior are given below.  In the near-field:
\ba
\label{eq:bare_self_energy_negative_plate_nearfield-long}
(4\pi/\coupling) \self^\infty(z)
&=&
-\frac{3}{4z}+\frac{1}{2}
- \frac{z}{9}-\frac{13 z^3}{675}
   +\frac{\left(-15 \sqrt{3} \ln\left(\left(2+\sqrt{3}\right) z\right)
   -15 \sqrt{3} \gamma +40\sqrt{3}+22\right)z^4}{2160}
   -\frac{2143z^5}{529200}
   \nonumber\\
   &+& \frac{\left(-105 \sqrt{3} \ln (z)-105 \sqrt{3} \gamma
   +357\sqrt{3}+250-105 \sqrt{3} \cosh ^{-1}(2)\right)z^6}{181440}
   -\frac{41323 z^7}{71442000}
   \nonumber\\
   &+& \frac{\left(-3150 \sqrt{3} \ln (z)-3150 \sqrt{3} \gamma +11905\sqrt{3}+5964-3150 \sqrt{3} \cosh^{-1}(2)\right)z^8}{36288000}
   + O(z^9).
\ea
In the \emph{far-field} region ($z \gg 1$), the correlation energy can be expanded into
\ba
(4\pi/\coupling) \self^\infty(z)
&=&
- \frac{3}{2}(2-\sqrt{3})  \ln (3)\, e^{-z}
+\left(
\frac{-323 + 188 \sqrt{3}}{4z} +
   3 (-7 + 4\sqrt{3}) (\gamma + \ln(108z))
   \right)\, e^{-2z}
   \nonumber\\
   &+& \left(\frac{3 \left(253
   \sqrt{3}-438\right)}{z}+\frac{3}{2} \left(15 \sqrt{3}-26\right) (8
   \ln (324 z)+8 \gamma +\ln (27))\right)e^{-3 z}
 \label{eq:self_energy_negative_plate_bare_farfield-long} \\
   &+& \left(\frac{4
   \left(3929 \sqrt{3}-6805\right)}{z}+6 \left(56 \sqrt{3}-97\right)
   (11 \ln (324 z)+11 \gamma -\ln (3))\right)e^{-4 z}
   + O(e^{-5 z}).  \nonumber
\ea
\end{widetext}

\end{document}